\documentclass{article}
\pdfoutput=1
\usepackage{jheppub}
\usepackage{amsmath}

\usepackage{tikz}

\newcommand{\roughly}[1]{\mathrel{\raise.3ex\hbox{$#1$\kern-0.85em\lower1ex\hbox{$\sim$}}}}

\newcommand{\lsim}{\roughly<}

\def\bfE{{\bf E}}
\def\bfJ{{\bf J}}

\def\bfn{{\bf n}}

\def\bfr{{\bf r}}

\def\bfx{{\bf x}}

\def\cA{{\cal A}}
 
\def\cC{{\cal C}}

\def\cL{{\cal L}}
\def\cM{{\cal M}}

\def\cO{{\cal O}}

\def\cS{{\cal S}}

\def\nn{\nonumber}
\def\({\left(}
\def\){\right)}
\def\[{\left[}
\def\]{\right]}

\newbox\charbox
\newbox\slabox
\def\slsh#1{{      
        \setbox\charbox=\hbox{$#1$}
        \setbox\slabox=\hbox{$/$}
        \dimen\charbox=\ht\slabox
        \advance\dimen\charbox by -\dp\slabox
        \advance\dimen\charbox by -\ht\charbox
        \advance\dimen\charbox by \dp\charbox
        \divide\dimen\charbox by 2
        \raise-\dimen\charbox\hbox to \wd\charbox{\hss/\hss}
        \llap{$#1$}
}}

\def\exd{{\hbox{d}}}

\def\nn{\nonumber}

\def\bea{\begin{eqnarray}}
\def\eea{\end{eqnarray}}
\def\be{\begin{equation}}
\def\ee{\end{equation}}

\def\ssB{{\scriptscriptstyle B}}

\def\ssN{{\scriptscriptstyle N}}

\def\KG{{\scriptscriptstyle KG}}

\def\Sch{{Schr\"odinger}~}

\def\pref#1{(\ref{#1})}

\title{Point-Particle Effective Field Theory II: Relativistic Effects and Coulomb/Inverse-Square Competition}

\author[a,b]{C.P.~Burgess,}
\author[a,b]{Peter Hayman,}
\author[a,b]{Markus Rummel,}
\author[c]{Matt Williams}
\author[a,b]{and L\'aszl\'o Zalav\'ari}
\affiliation[a]{Physics \& Astronomy, McMaster University, Hamilton, ON, Canada, L8S 4M1}
\affiliation[b]{Perimeter Institute for Theoretical Physics, Waterloo, Ontario N2L 2Y5, Canada }
\affiliation[c]{Instituut voor Theoretische Fysica, KU Leuven,
Celestijnenlaan 200D,
B-3001 Leuven, Belgium}

\date{\today}

\abstract {We apply point-particle effective field theory (PPEFT) to compute the leading shifts due to finite-sized source effects in the Coulomb bound energy levels of a relativistic spinless charged particle. This is the analogue for spinless electrons of calculating the contribution of the charge-radius of the source to these levels, and our calculation disagrees with standard calculations in several ways. Most notably we find there are {\em two} effective interactions with the same dimension that contribute to leading order in the nuclear size, one of which captures the standard charge-radius contribution. The other effective operator is a contact interaction whose leading contribution to $\delta E$ arises {\em linearly} (rather than quadratically) in the small length scale, $\epsilon$, characterizing the finite-size effects, and is suppressed by $(Z\alpha)^5$. We argue that standard calculations miss the contributions of this second operator because they err in their choice of boundary conditions at the source for the wave-function of the orbiting particle. PPEFT predicts how this boundary condition depends on the source's charge radius, as well as on the orbiting particle's mass. Its contribution turns out to be crucial if the charge radius satisfies $\epsilon \lsim (Z\alpha)^2 a_\ssB$, where $a_\ssB$ is the Bohr radius, because then relativistic effects become important for the boundary condition. We show how the problem is equivalent to solving the Schr\"odinger equation with competing Coulomb, inverse-square and delta-function potentials, which we solve explicitly. A similar enhancement is {\em not} predicted for the hyperfine structure, due to its spin-dependence. We show how the charge-radius effectively runs due to classical renormalization effects, and why the resulting RG flow is central to predicting the size of the energy shifts (and is responsible for its being linear in the source size). We discuss how this flow is relevant to systems having much larger-than-geometric cross sections, such as those with large scattering lengths and perhaps also catalysis of reactions through scattering with monopoles. Experimental observation of these effects would require more precise measurement of energy levels for mesonic atoms than are now possible.}

\begin{document}
\maketitle
\section{Introduction}
\label{section:intro}

Effective field theories (EFTs) provide an efficient means to analyse systems that involve a large hierarchy of scales, in applications for which only the longer of the scales (in distance or time) is to be directly probed. We here extend results from a companion paper \cite{SchInvSq} to some issues arising from the application of EFTs to a particular system of this type: the problem of a spinless point charge --- which we henceforth generically call a `meson' or `spinless-electron' interacting electromagnetically with another charged particle that might contain some substructure (like a proton or a nucleus -- henceforth called the `nucleus' or `source'). 

In this case the hierarchy of interest is the ratio between the large size, $a$, of the meson orbit (or point of closest approach), relative to the `nuclear' size, $\varepsilon$. Because it is specifically the ratio $\varepsilon/a$ that we wish to follow, we simplify the discussion by taking the limit where the nuclear mass is infinitely large, $M/m \to \infty$. This allows us to focus more efficiently on finite-size issues within the context of motion within a fixed Coulomb field. We argue that surprises potentially lurk even in this restricted regime.

When $\varepsilon \ll a$ we expect an appropriate effective description to lead efficiently to a series expansion of observables in powers of $\varepsilon/a$. Effective theories capture this expansion by writing an effective action for a source with structure, which includes all possible interactions involving the `bulk' fields of interest (like the EM field $A_\mu$ or meson field $\psi$, say) consistent with the symmetries of the problem. In practice one organizes these interactions with increasing (mass-) dimension\footnote{We use fundamental units for which $\hbar = c = 1$.} in the expectation that dimensional analysis then requires the couplings of higher-dimension interactions to be suppressed by additional powers of $\varepsilon$ relative to couplings of lower-dimensional interactions. 

Of course, simply writing down a point-particle action is not new in itself. The new part --- and what we mean by `point-particle effective field theory' (or PPEFT) --- is the explicit connection that is made between this action and the near-source boundary conditions for the various `bulk' fields to which it couples (this connection is laid out more formally in \cite{SchInvSq}, building on the earlier construction of \cite{EFTCod2}). It is through these boundary conditions that the effective couplings of the source action can influence the integration constants arising when solving bulk field equations, and thereby express how the source back-reacts onto its surrounding environment.

Concretely, for a rotationally invariant nucleus coupled to photons and spinless electrons (respectively described by the bulk fields $A_\mu$ and $\psi$), such an effective action might have lowest-dimension interactions of the form
\be \label{sourceaction}
  S_b = -\int \exd \tau \; \Bigl[ M - Q \, A_\mu \dot y^\mu +  h\, \psi^* \psi  - {\tilde h} \, \nabla \cdot \bfE + \cdots \Bigr] \,,
\ee
where the integral is along the world-line, $y^\mu(\tau)$, of the nucleus for which $\tau$ is the proper time and $\dot y^\mu := \exd y^\mu/\exd \tau$. 

The constants $M$, $Q$ represent the nuclear mass and charge (we take $Q=Ze$), while the couplings $h$ and ${\tilde h}$ are the first of a succession of possible effective couplings having dimensions that are a positive power of length. The rest of these terms are collectively denoted by the ellipses in \pref{sourceaction}, and include all possible local interactions involving $A_\mu$ and $\psi$ and their derivatives, and it turns out that all of those not written are negligible for the present purposes because they are suppressed by more powers of the small scale $\varepsilon$ than are those explicitly written.\footnote{The dimension of the interaction depends on the canonical dimension of $\psi$, which is mass for a Klein-Gordon field but mass${}^{3/2}$ for a Schr\"odinger field. In most of what follows it is the Schr\"odinger field that is of interest, though we switch to a Klein-Gordon field for the discussion of relativistic effects below.}  In what follows we keep only the above three terms, dropping all other effective interactions with higher mass dimensions than these.

The coupling ${\tilde h}$ describes the traditional charge-radius of the nucleus. It is related to the root-mean-square charge radius, $r_p^2$, by ${\tilde h} = \frac16\, Ze \, r_p^2$, as might be measured by scattering photons from the nucleus. For observables not involving photons the electromagnetic field may be integrated out, which amounts in this case to using Maxwell's equations to rewrite $\nabla \cdot \bfE$ in terms of the total charge density, which for a Schr\"odinger field is $\rho = - e \,\psi^* \psi + \rho_\ssN$, where $\rho_\ssN = Ze \left[ 1 + \frac16 \, r_p^2 \, \nabla^2 + \cdots \right] \delta^3(\bfr)$ is the rest-frame nuclear charge density obtained by varying $S_b$ with respect to $A_0$. The term quadratic in $\psi$ can then be absorbed into $h$, leading to an effective interaction of the form $- h_{\rm tot} \, \psi^* \psi$ with  
\be \label{htot}
  h_{\rm tot} \simeq h + \frac16 \, Z e^2 \, r_p^2 \,.
\ee

A naive estimate for how $- h_{\rm tot} \, \psi^*\psi$ contributes to physical observables comes from recognizing that it is equivalent to a delta-function potential of the form $\delta V = h_{\rm tot} \, \delta^3(\bfr)$, and indeed using $h_{\rm tot}  = \frac16 \, Z e^2 r_p^2$ in the perturbative formula $\delta E = h_{\rm tot} |\psi(0)|^2$ using standard Coulomb wave-functions reproduces the leading expression for the nuclear-radius contribution to atomic energy shifts \cite{CODATA}.

If $\psi$ had been a Klein-Gordon scalar then $h = h_\KG$ would have dimension length (rather than length-squared) and so naively might be expected to contribute to observables linearly in the small scale $\varepsilon$. The main point of this paper is to argue that this is basically true for {\em both} Schr\"odinger and Klein-Gordon fields. (If $\psi$ is a Schr\"odinger field, it turns out only the combination $h_\KG = 2mh$ contributes to observables and it is this combination that scales linearly in $\varepsilon$.) It is also more subtle than it looks even for Klein-Gordon fields.  

Two surprises turn out to be buried within the statement that $h$ scales linearly with $\varepsilon$:
\begin{itemize}
\item {\em Reaction `catalysis':} Although the leading influence of $h$ on physical observables is linear in microscopic scales, it turns out that the scale involved need not strictly be $\varepsilon$ and in some cases can be much larger. In particular, because we find that the coupling $h$ must be renormalized --- {\em even at the classical level} --- it runs with scale according to a renormalization-group (RG) evolution. It therefore contributes to observables proportional to the RG-invariant scale, $\epsilon_\star$, associated with this running, which can (but need not) be much larger than the underlying physical scale $\varepsilon$. When $\epsilon_\star \gg \varepsilon$ physical processes like scattering can be strongly enhanced, in a way that resembles how scattering from magnetic monopoles can catalyze \cite{monopolecatal, monopolerev} the violation of baryon number in grand-unified theories.\footnote{We argue this running is a part of the mechanism for understanding monopole catalysed events within the PPEFT.} 
\item {\em Larger than expected shifts in atomic energy levels:} Even when $\epsilon_\star \simeq \varepsilon$ we argue that $h$ shift energy levels (and affects scattering) in surprising ways. First, because (for the Schr\"odinger field) only the combination $h_\KG = 2mh$ appears in physical quantities, when the orbiting particle is relativistic at nuclear radii then matching to a nucleus leads to the expectation $h = B/m$ where $B$ is of nuclear size and independent of the $\psi$ mass. (The same need not be true when it is nonrelativistic at the nuclear surface.) This leads to unexpected shifts in the energy levels of spinless particles that are of order $\delta E \propto h \propto \varepsilon/m$. Beyond this, the classical renormalization adds additional $m$-dependence in the influence of $h$ on observables. In particular, although $h = 0$ is an RG fixed point for a non-relativistic particle experiencing only a Coulomb potential, it is {\em not} a fixed point for a relativistic particle in a Coulomb potential or for a non-relativistic particle experiencing a superposition of both Coulomb and inverse-square potentials. Because zero coupling is not in this case a fixed point, contact interactions become compulsory rather than optional: $h=0$ can at best only be chosen at a particular scale (perhaps at the UV scale $\varepsilon$). If so then $h$ runs to become nonzero at larger scales and where it contributes to observables linearly in $\epsilon_\star$. In particular, for hydrogen-like states both of these effects imply $s$-wave states are shifted in energy by amounts that depend differently on mass than does the normal $(Z\alpha)^4r_p^2m^3$ charge-radius term (where $\alpha$ is the usual fine-structure constant). Unfortunately\footnote{`Unfortunately' because if shared by spin-half particles such effects have the right size and sign to have accounted for the experimental `proton-radius' discrepancy \cite{ProtonRadiusSummary, ProtonRadiusStandards, Contact, MuonLamb, CODATA} without the need for exotic new interactions \cite{ProtonRadiusNewPhys}. } these effects seem not to be shared by spin-half pariticles \cite{Dirac}, and so their experimental verification requires more precise measurements for the energies of $\pi$- or $K$-mesic atoms than are presently possible.
\end{itemize}

\subsection*{Contact interactions, boundary conditions and classical renormalization}

Although we fill in the details explicitly in the bulk of the paper, because the results are so surprising we first provide here a brief sketch of the logic of the argument. 

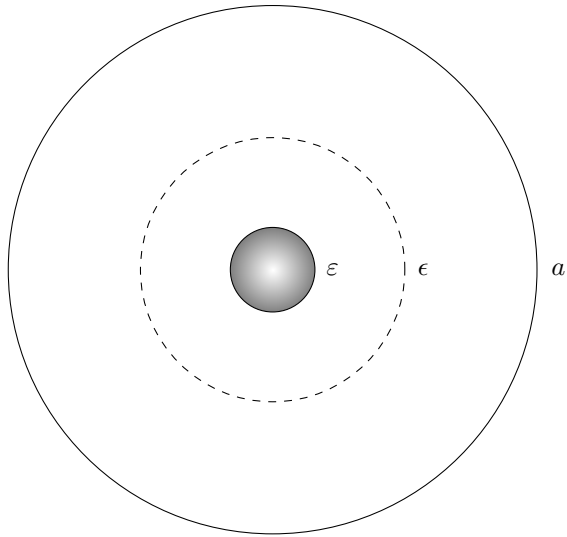
\begin{figure}
	\centering
\begin{tikzpicture}
	\shadedraw[inner color=white, outer color=gray, draw=black] (0,0) circle [radius=16pt];
	\draw (0.8,0) node {$\varepsilon$};
	\draw[dashed] (0,0) circle [radius=50pt];
	\draw (2.0,0) node {$\epsilon$};
	\draw (0,0) circle [radius=100pt];
	\draw (3.8,0) node {$a$};
\end{tikzpicture}
\caption{A schematic of the scales arising in the boundary conditions near the source. We denote by $\varepsilon$ the actual UV physics scale associated with the underlying size of the source (e.g., the size of the proton), which by assumption is very small compared to the scale $a$ of physical interest (e.g., the size of an atom). The PPEFT uses the action of the point source to set up boundary conditions on the surface of a Gaussian pillbox of radius $\epsilon$. The precise size of this pillbox is arbitrary, so long as it satisfies $\varepsilon \ll \epsilon \ll a$. We require $\varepsilon \ll \epsilon$ in order to have the first few multipole moments (in our example only the first is considered) dominate the field on the surface of the pillbox, and we require $\epsilon \ll a$ in order to be able to truncate the effective action at the few lowest-dimension terms. The classical RG flow describes how the effective couplings within the PPEFT action must change for different choices of $\epsilon$ in order to keep physical quantities unchanged.} \label{Figcirc}
\end{figure}

The crucial role is played by the coupling $h$, of the lagrangian \pref{sourceaction}, which from the point of view of the $\psi$ field equation appears as would a `contact' interaction ({\em i.e.} a delta-function contribution to the inter-particle interaction potential).\footnote{Such contact interactions sometimes arise in the Coulomb problem, such as to describe strong meson-nucleus interactions in mesonic atoms where a negative pion or kaon orbits the nucleus.} As might be expected, even in the absence of Coulomb interactions, the presence of a delta-function potential necessarily modifies the boundary condition that $\psi$ satisfies at the origin, as can be seen by integrating the field equations over an infinitesimal Gaussian pillbox that encircles the source nucleus (see Fig.~\ref{Figcirc} and the discussion in \cite{SchInvSq}). This implies that as $r \to 0$ the radial derivative satisfies\footnote{If one resists imposing this boundary condition, such as by perturbing in the interaction $h$, one finds that graphs involving repeated meson interactions with the nucleus are not small and their resummation \cite{BraneRenorm} simply imposes \pref{newbc}.}
\be \label{newbc}
  4\pi r^2 \frac{\partial \psi}{\partial r} = \lambda \psi \,, 
\ee
where $\lambda = h_\KG = 2mh$. It is $\lambda$, rather than $h$, that is approximately independent of $m$ for sources small enough that orbiting particles are relativistic in their vicinity, a feature we further motivate in the Appendix using several toy models of the nuclear charge distribution.

Recall that for free particles (or for particles interacting through the Coulomb interaction) the two independent solutions to the radial equation behave for small $r$ like $\psi_+ \sim r^\ell$ and $\psi_- \sim r^{-\ell-1}$, for angular-momentum quantum number $\ell$. When $h = 0$ eq.~\pref{newbc} reduces to the usual condition that the overlap with $\psi_-$ must vanish. More generally the solution satisfying \pref{newbc} involves both $\psi_+$ and $\psi_-$. But because $\psi_-$ diverges\footnote{Regularity at the origin is not in itself a good boundary condition for two reasons. First, it is generic that bulk fields diverge at the position of a source -- as is clearest for the Coulomb potential, $A_0 \propto q/(4\pi r)$. Second, for many simple potentials (such as the inverse-square: $V \propto 1/r^2$) it can happen that {\em both} radial solutions diverge at $r = 0$, so boundedness cannot distinguish them. } as $r \to 0$ the use of \pref{newbc} requires it to be regulated and evaluated at infinitesimal $r = \epsilon$ rather than strictly at zero. 

Once this is done \pref{newbc} makes sense, but also seems to require that physical quantities must depend in detail on the value of the regularization scale $\epsilon$, which seems odd given we at this point only needed to choose $\epsilon$ to be small and not precisely equal to the physical UV scale $\varepsilon$. What really happens though is that physical quantities are $\epsilon$-independent because the explicit $\epsilon$'s in \pref{newbc} can be renormalized into $h$ \cite{Jackiw}. That is, the explicit $\epsilon$-dependence of \pref{newbc} can cancel against an implicit $\epsilon$-dependence buried in $h(\epsilon)$, which turns out to require $h = \epsilon \, f(h_0/\epsilon_0,\epsilon/\epsilon_0)$ where $f(x,y)$ is a nontrivial dimensionless function (given explicitly below) and $\epsilon_0$ is a  scale where $h = h_0$. This required $\epsilon$-dependence of $h$ is what we call its renormalization-group (RG) evolution. An RG invariant scale $\epsilon_\star$ can then be defined, such as by specifying the scale where $h_0 = 0$ or $h_0 \to \infty$. In the end it is only $\epsilon_\star$ on which physical quantities typically {\em do} depend, and this is ultimately the origin of the first bullet point given above.

A reality check on this running is that it has a fixed point at $h=0$, in the sense that $f(0,y) = 0$ for all $y$. This means (for delta-functions plus Coulomb potentials, at least) one can always choose not to have a contact interaction if that is what one wants. That is, once $h = 0$ at any scale, then its running ensures it remains zero for all scales. And if $h$ is nonzero at some scales it turns out that its flow is towards the fixed point at zero in the far infrared (IR), as $\epsilon \to \infty$.

\subsection*{Energy shifts}

So far so good. Where the real surprises start is once the Coulomb and delta-function interactions are supplemented by an attractive inverse-square potential:
\be \label{cartoonpot}
 V = - \frac{s}{r} - \frac{g}{r^2} + h \, \delta^3(\bfx) \,.
\ee
for nonzero and positive $g$. The starting observation is that such an inverse-square potential can compete with the centrifugal barrier and so modify the asymptotic form of solutions as $r \to 0$ to become more singular there.\footnote{When the boundary condition is imposed at finite $r = \epsilon$ rather than zero, the presence of an inverse-square interaction competing with the Coulomb potential can be expected to be important if $\epsilon \lsim g/s$.} As many have observed \cite{EssinGriffiths, InvSqPot, Efimov, Kaplan} because of this the Schr\"odinger Hamiltonian can fail to be self-adjoint, depending the boundary conditions that hold at $r = \epsilon$. Selecting a choice of boundary condition to secure its self-adjointness --- not a unique construction --- is known as constructing its self-adjoint extension \cite{SAext,SAext2}. 

We here follow closely the treatment of inverse-square potentials given in \cite{SchInvSq}. Physically, what is happening is that the inverse-square potential concentrates $\psi$ more at the origin and so increases the probability of interacting with whatever the physics is that describes the source there. In particular the resulting time-evolution could be non-unitary if these interactions excite other degrees of freedom besides $\psi$ (or otherwise remove $\psi$ particles for whatever reason). On the other hand interactions with the source might preserve probability if there are no other degrees of freedom and the action describing the source is real. In this language the freedom inherent in choosing self-adjoint extensions is precisely the freedom in choosing the form of the source action. In particular, given an action like \pref{sourceaction} the boundary condition \pref{newbc} specifies a specific self-adjoint extension provided $h$ is real. But casting the extension in terms of the source action has the advantage that it gives a criterion for how to choose it; in particular it shows why the lowest-dimension interactions should dominate at low energies, and so why \pref{sourceaction} should commonly apply at low-energies.\footnote{Because quadratic actions often dominate, the RG evolution described here is likely universal for many systems.}

In detail, the way this connection between inverse-square potentials and contact interactions arises is through the contribution the inverse-square potential makes to the RG evolution of $h$. (This observation is also not in itself new, since it has long been known to be an example of renormalization and dimensional transmutation within a quantum mechanical setting.\footnote{ When cast in terms of a self-adjoint extension it has not always been clear --- see however \cite{Kaplan} --- that it is usually the strength of a delta-function contact potential that is being renormalized.}) Naively, perturbing in $h$ leads to formulae like $\delta E = h |\psi(0)|^2$, however the inverse-square potential causes $\psi(0)$ to diverge and this turns the expansion in $h$ into a more dangerous expansion in $h/\epsilon$ that breaks down as $\epsilon \to 0$. It is this nontrivial dependence on $h/\epsilon$ that the RG efficiently resums. 

What is important, however, is that when $g \ne 0$ the IR fixed point of the RG evolution for $h$ gets driven away from $h = 0$ towards a nonzero value. As a result the presence of a contact interaction becomes compulsory, rather than merely being an option. At best $h = 0$ can only hold at a specific scale, $\epsilon_0$ say, after which RG evolution requires it not to vanish anywhere else. This is equally true if a Coulomb interaction is also present or not. 
Contributions of the contact interaction to bound-state energy levels and scattering amplitudes turn out to be linear in $\epsilon_\star$ or $\epsilon_0$, and so contribute negligibly if these scales should be vanishingly small, as would be true if the value of $h$ were set at a vanishingly small length scale (as appropriate if the nucleus were a point particle like a muon).

Most crucially, an inverse-square potential is {\em always} present for the relativistic Coulomb problem. We consider here the Klein-Gordon/Coulomb system, for which the square of the Coulomb potential appears within the second time derivative,\footnote{Field redefinitions allow effects to be moved around within an EFT \cite{EFTrev}, and if the lowest-order Schr\"odinger equation is used to eliminate $D_t ^2\psi$ the self-adjointness problems remain, being attributable now to the appearance of higher spatial derivatives like $\nabla^4 \psi$ in the action. } $D_t^2 \psi$. The radial part of the Klein-Gordon/Coulomb equation is precisely the same as in \pref{cartoonpot} with $s \simeq 2m Z \alpha$ while $g = (Z\alpha)^2$. From the above discussion, the presence of the relativistic inverse-square potential ensures $h = 0$ is not a fixed point, provided boundary conditions are required at distances as small as $\epsilon \lsim g/s \simeq Z\alpha/m$. When this is so $h$ must instead be driven away from zero in the far infrared to be of order $2m h \sim \epsilon_\star$. 

The presence of such a contact interaction then turns out to shift $s$-wave energy levels by an amount $\delta E \sim (Z \alpha )^3 \epsilon_\star m^2$, that is linear in $\epsilon_\star$ as claimed in the second bullet point above. Linearity in the microscopic scale $\epsilon_\star$ is unlike standard contributions to energy shifts due to nuclear finite-size effects \cite{FiniteSize}. It is intriguing that for plausible nuclear values of $\epsilon_\star$ this energy shift is similar to what is seen experimentally in comparisons between energy levels for electrons and muons bound to nuclei, though (alas) similar-sized contributions do not also arise in the Dirac equation appropriate for spin-half particles \cite{Dirac}. 

\subsubsection*{A road map}

The rest of this paper is organized as follows. The next section, \S\ref{sec:NRmixedCISP}, sets up and solves the Schr\"odinger equation in the presence of the potential \pref{cartoonpot}. Both bound-state energies and scattering amplitudes are computed explicitly, as is the detailed RG flows for the contact coupling $h$ for several regimes that differ according to the size of the inverse-square coupling $g$. \S\ref{sec:Apps} then follows this with several applications of these results, designed as checks or illustrations of the two bullet points given above. A reality check first uses the results of \S\ref{sec:NRmixedCISP} to derive the Deser formula \cite{Deser} for mesonic atoms, that relates the energy-level shift of $s$-wave states and the low-energy scattering length due to the short-range meson-nuclear force. This is followed by a short discussion identifying under what circumstances the RG evolution of the contact interaction can enhance scattering cross sections, as is reminiscent of monopole-catalyzation of exotic GUT-scale reactions. Next is a detailed treatment of the Klein-Gordon/Coulomb system, and the estimates of the size of the energy-level shifts that are implied by the running of the contact interaction. Some toy models checking these results against specific nuclear charge distributions are also considered in the Appendix, and are compared with the more general EFT estimates.

\section{Nonrelativistic mixed Coulomb and inverse-square potentials} 
\label{sec:NRmixedCISP}

Much of the physics needed for the relativistic case hinges on the competition between the inverse-square and Coulomb potentials, so we start our discussion with the Schr\"odinger system involving these two potentials. Our treatment follows closely that of \cite{SchInvSq}, which examines the classical renormalizations associated with the inverse-square potential, though we extend this analysis here by adding also a Coulomb potential.  

\subsection{Schr\"odinger action}

We take, therefore, our action to be $S = S_\ssB + S_b$ where $S_\ssB$ is the Schr\"odinger `bulk' action 
\be \label{Sbulk}
 S_\ssB = \int \exd t\, \exd^3 x \left\{ \frac{i}2 \, \Bigl(\Psi^* \partial_t \Psi -  \Psi \, \partial_t \Psi^* \Bigr) - \Psi^* \left[ - \frac{1}{2m} \, \nabla^2 + V(\bfx) \right] \Psi \right\} \,,
\ee
where $m$ is the particle mass and $S_b$ describes a microscopic contact interaction between the Schr\"odinger field and the point source localized at the origin $r=0$:
\be \label{Ssource}
 S_b = \int \exd t\, \cL_b[ \Psi(x=0), \Psi^*(x=0)] = \int \exd t\, \exd^3 x \, \cL_b(\Psi, \Psi^*) \, \delta^3(\bfx) \,,
\ee
with 
\be \label{VLb}
  V(\bfx) = - \frac{s}{r} - \frac{g}{r^2} \quad \hbox{and} \quad
  \cL_b = -h \, \Psi^* \Psi 
\ee
used, for coupling constants $s$, $g$ and $h$, when an explicit form is required. 

The field equation found by varying $\Psi^*$ then is the Schr\"odinger equation,
\be
 i \partial_t \Psi = - \frac{1}{2m} \, \nabla^2 \Psi + V(\bfx) \Psi + \frac{\partial \cL_b}{\partial \Psi^*} \, \delta^3(\bfx)  \,,
\ee
which for energy eigenstates, $\Psi(\bfx,t) = \psi(\bfx) \, e^{-iEt}$, and with the choice \pref{VLb} becomes
\be \label{SchE}
  \nabla^2 \psi + 2m \left[ \frac{s}{r} + \frac{g}{r^2} - h \, \delta^3(\bfx) \right] \psi = \kappa^2 \psi \,,
\ee
with $\kappa^2  = - 2mE$. For bound states --- when $E \le 0$ --- $\kappa$ is real, but when discussing scattering --- where $E \ge 0$ --- we switch to $\kappa = ik$ with real $k$ given by $k^2 = + 2mE$.  Expanding in spherical harmonics, $Y_{\ell \ell_z}(\theta,\phi)$, implies the radial equation is given by
\be \label{SchEr}
  \frac{1}{r^2} \frac{\exd}{\exd r} \left( r^2 \frac{\exd \psi_{\ell \ell_z}}{\exd r} \right) - \left[ \frac{\ell(\ell+1)}{r^2} + U (r) \right] \psi_{\ell \ell_z} 
 = \kappa^2 \, \psi_{\ell \ell_z}  \,,
\ee
where $U = 2m[V + h \, \delta^3(\bfx)]$ while $\ell = 0,1,2,\cdots$ and $\ell_z = -\ell, -\ell +1,\cdots,\ell-1,\ell$ are the usual angular momentum quantum numbers.

\subsection{Source action and boundary conditions}

The source action, $S_b$, appears here only through the delta-function contribution to $U$ and the only effect of this is to determine the boundary condition satisfied by $\psi$ at $r = 0$. This can be obtained as described in \cite{SchInvSq} by integrating \pref{SchE} over an infinitesimal sphere, $\cS$, of radius $0 \le r \le \epsilon$ around $\bfx = 0$ and using continuity of $\psi$ there to see that only the integral of the second derivative contributes from the left-hand side of \pref{SchE} as $\epsilon \to 0$. This leads to the result
\be
 \lambda \, \psi(0) = \int_\cS \exd^3 x \, \nabla^2 \psi =  \int_{\partial \cS} \exd^2 x \, \bfn \cdot  \nabla \psi =   \int \exd^2 \Omega \, \left( r^2 \,\frac{\partial \psi}{\partial r} \right)_{r = \epsilon} = 4\pi \epsilon^2 \left( \frac{\partial \psi}{\partial r} \right)_{r = \epsilon} \,,
\ee
where $\lambda := 2mh$ while $\bfn \cdot \exd \bfx = \exd r$ is the outward-pointing radial unit vector, $\exd^2 \Omega = \sin \theta\, \exd \theta\, \exd \phi$ is the volume element on the surface of the angular 2-sphere and the last equality assumes a spherically symmetric source so that $\psi$ is also spherically symmetric to good approximation for $\epsilon$ sufficiently small.

Because solutions $\psi_{\ell m}(r)$ vary like a power $r^p$ as $r \to 0$, the boundary condition given above becomes singular as $\epsilon \to 0$. This is dealt with by renormalizing $\lambda$ --- {\em i.e.} by associating an implicit $\epsilon$-dependence to $\lambda$ in such a way as to ensure that the precise value of $\epsilon$ drops out of physical predictions. With this in mind --- and defining $\psi(0) := \psi(r=\epsilon)$ --- our problem is to solve the radial equation, \pref{SchEr}, subject to the boundary condition 
\be \label{lambdabc}
  \left[4\pi r^2 \,  \frac{ \partial}{\partial r} \ln \psi \right]_{r=\epsilon}  =  \lambda \,,
\ee
at the regulated radius $r = \epsilon$. 

As mentioned in \cite{SchInvSq}, this boundary condition can be regarded as a specific choice of self-adjoint extension \cite{SAext,SAext2} of the inverse-square Hamiltonian. The inverse-square potential requires such an extension because its wave-functions are sufficiently bunched at the origin that physical quantities actually care about the nature of the physics encapsulated by the source action, $S_b$. Writing the extension in this way usefully casts its ambiguities in terms of a physical action describing the physics that can act as a potential sink (or not) of probability at $r = 0$. As might be expected, this extension is self-adjoint provided that the source action is real and involves no new degrees of freedom. In the present instance this can be seen from the radial probability flux, 
\be
 J = 4\pi r^2 \, \bfn \cdot \bfJ = \frac{2\pi r^2}{m} \Bigl(  \Psi \partial_r \Psi^* - \Psi^* \partial_r \Psi  \Bigr) \,,
\ee
emerging from the source through the surface at $r = \epsilon$. Evaluating with energy eigenstates gives
\be
 J(\epsilon) = \frac{2\pi \epsilon^2}{m} \Bigl[ \psi(\epsilon)\partial_r \psi^*(\epsilon) - \psi^*(\epsilon) \partial_r \psi(\epsilon)\Bigr] =  (h^* - h) \;\psi^* \psi(\epsilon) \,,
\ee
which shows no probability flows into or out of the source when its action is real (ie $h^* = h$).

\subsection{Solutions}

The radial equation \pref{SchEr} to be solved is
\be \label{SchEgk}
   r^2 \frac{\exd^2 \psi}{\exd r^2} + 2r \, \frac{\exd \psi}{\exd r} + \left( wr + v - \kappa^2 r^2 \right) \psi
 = 0\,,
\ee
where $w = 2ms$ and $v = 2mg - \ell(\ell+1)$. This can be written in confluent hypergeometric form through the transformation $\psi(r) = z^l \, e^{-z/2} u(z)$, for $z = 2\kappa r$ where $l(l+1) + v = 0$ so that\footnote{Choosing the other root for $p$ just exchanges the roles of the two independent solutions encountered below, so does not introduce any new alternatives.} 
\be \label{ldef}
  l = \frac12 ( -1 + \sqrt{1 - 4 v} ) = \frac12 ( -1 + \zeta ) = - \frac12 + \sqrt{ \left( \ell + \frac12 \right)^2 - \xi} \,,
\ee
where we define for later notational simplicity $\xi := 2mg$ and
\be \label{zetadef}
 \zeta := \sqrt{1 - 4v} = \sqrt{1 + 4 \ell(\ell+1) - 4\xi} = \sqrt{(2\ell+1)^2 - 4 \xi}\,.
\ee
The two linearly independent radial profiles therefore are 
\be \label{psipm}
 \psi_\pm(r)  = (2\kappa r)^{\frac12(-1\pm \zeta)} \, e^{-\kappa r}  \cM\left[ \frac12 \left(- \frac{w}{\kappa} + 1 \pm \zeta\right), 1 \pm \zeta ; 2\kappa r \right] \,,
\ee
where $\cM(a,b;z) = 1 + (az/b) + \cdots$ is the confluent hypergeometric function regular at $z = 0$. We therefore take our general radial solution to have the form $\psi = C_+ \psi_+ + C_- \psi_-$.

We next impose the boundary condition at $r = 0$ to determine the ratio $C_-/C_+$. Regularizing for small $r = \epsilon$ the solutions $\psi_\pm(r)$ behave as
\be
 \psi_\pm(\epsilon) = (2 \kappa \epsilon)^{\frac12(-1\pm \zeta)} \left[ 1 - \frac{w \epsilon}{1 \pm \zeta} + \cO(\epsilon^2) \right] \,,
\ee
which has the familiar form of $r^l$ or $r^{-l-1}$, with $l$ as defined in \pref{ldef}. This shows that for some choices of $\xi$ {\em neither} of $\psi_\pm$ is bounded at the origin. This implies that boundedness at the origin cannot be the right physical criterion there, at least in the presence there of a physical source. This is not really a surprise since fields generically diverge at the presence of a source, such as does the Coulomb potential itself.  

We do demand solutions be normalizable, however, and the convergence of the integral $\int \exd^3 x \, |\psi|^2$ as $r \to 0$ implies $\psi$ cannot diverge faster than $r^{-3/2}$ as $r \to 0$. For $\psi_\pm$ this implies $2 \pm \zeta > 0$. For concreteness' sake in what follows we follow \cite{SchInvSq} and specialize to the case where the inverse-square potential satisfies $-\frac 34 < \xi < \frac54$, because this captures all of the examples of most interest and has the property that $\psi_-$ is not normalizable at $r = 0$ for any $\ell \ne 0$. This ensures that that the boundary condition at the origin implies $C_- = 0$ and so $\psi \propto \psi_+$ for $\ell \ne 0$.

It is only for $\ell = 0$ that the contact interaction is needed to determine $C_-/C_+$, and for such $s$-wave states we have $\zeta(\ell = 0) = \zeta_s := \sqrt{1 - 4\xi}$ and so $0 \le \zeta < 1$ for $0 \le \xi \le \frac14$, and so both solutions diverge but are normalizable at the origin.\footnote{The only exception to this is the case $\xi = 0$ for which $l = \ell$ and so $\psi_+$ is bounded. However once having discarded boundedness as a valid criterion at the origin, it cannot be revived in this special case. In our view this is a deficiency of most treatments of the Coulomb potential, a point to which we return below.} If $\frac14 < \xi \le \frac54$ then $\zeta_s$ becomes imaginary, in which case both $|\psi_+|^2$ and $|\psi_-|^2$ diverge near $r = 0$ while remaining normalizable. In this case eq.~\pref{lambdabc} is the condition that fixes $C_-/C_+$, evaluating the derivative using the small-$r$ form for $\psi_\pm$ leads to 
\bea \label{bceps}
 \lambda = 4\pi \epsilon^2 \left( \frac{\partial }{\partial r} \,\ln \psi \right)_{r=\epsilon} &=& 2\pi \kappa \epsilon^2 \left[ \frac{C_+ \left( -1 + \zeta_s \right) (2\kappa \epsilon)^{\frac12(-3+\zeta_s)} + C_- \left( -1 - \zeta_s \right) (2\kappa \epsilon)^{\frac12(-3-\zeta_s)}}{ C_+  (2\kappa \epsilon)^{\frac12(-1+\zeta_s)} + C_-   (2\kappa \epsilon)^{\frac12(-1-\zeta_s)}} \right] \nn\\
 &=& -2\pi \epsilon \left[ 1 + \zeta_s \left( \frac{R-1}{R+1}\right)  \right]  \,,
\eea
where
\be \label{Rexp}
 R := \left( \frac{C_-}{C_+} \right) \, (2\kappa \epsilon)^{-\zeta_s} \,,
\ee
and so, in particular, $R=0$ when $C_- = 0$.

To use this equation it is useful to rewrite it as
\be \label{bceps2}
\hat \lambda := \frac{\lambda}{2\pi \epsilon}  + 1 =   \zeta_s \left( \frac{1-R}{1+R} \right) \,,
\ee
where the first equality defines\footnote{Notice that vanishing coupling, $\lambda = 0$, corresponds to $\hat \lambda = 1$, and so attractive (repulsive) $\delta$-potentials corresponding to $\lambda < 0$ ($\lambda > 0$) imply $\hat\lambda < 1$ ($\hat\lambda > 1$). } the dimensionless coupling $\hat\lambda$. This shows that physical quantities depend only on the ratio $\hat \lambda/\zeta_s$. Solving for $C_-/C_+$ leads to
\be \label{Rvslamzet}
 \frac{C_-}{C_+} = R(\hat\lambda/\zeta_s) (2\kappa \epsilon)^{\zeta_s} = \left[ \frac{\zeta_s - \hat \lambda}{\zeta_s + \hat \lambda} \right] (2\kappa \epsilon)^{\zeta_s} \,.
\ee
This is positive if $|\hat \lambda| \le \zeta_s$ and negative otherwise. For scattering calculations we take $\kappa = ik$ and then \pref{Rvslamzet} fixes $\psi(r)$ up to normalization, thereby allowing scattering phases to be read off by examining the large-$r$ limit. Alternatively, for bound states it is the compatibility of \pref{Rvslamzet} with the value $C_-/C_+$ obtained by the normalization condition at infinity that picks out the quantized value for $\kappa$ (and so also $E = - \kappa^2/2m$). 

Two points about this boundary condition are noteworthy:
\begin{itemize}
\item Even though $\zeta_s$ need not always be real \pref{bceps2} always amounts to a single real condition on $C_-/C_+$ or $\kappa$, because $R$ is either real (when $\zeta_s$ is real, and so $v < \frac14$) or $R$ is a pure phase (when $\zeta_s$ is pure imaginary, and so when $v > \frac14$). Our main interest is in small $v$, so in what follows we restrict attention to real $\zeta_s$.
\item Although \pref{bceps2} seems to imply $\kappa$ depends on $\epsilon$, this naive dependence is cancelled by the $\epsilon$-dependence implicit in the renormalization of $\lambda$. The required $\epsilon$-dependence is worked out below separately for the two cases where $\zeta_s$ is real or imaginary.
\end{itemize}

\subsection{RG evolution}

The $\epsilon$-dependence of $\lambda$ required to make physical quantities like $\kappa$ independent of $\epsilon$ can be found by differentiating the quantization condition \pref{bceps2} or \pref{Rvslamzet}, being careful to hold physical quantities like $\kappa$ or $C_-/C_+$ fixed. We focus here on real $\zeta_s$, though the imaginary case goes through along the lines found in \cite{SchInvSq} since the RG discussion does not depend on the Coulomb interaction.

When $\zeta_s$ is real then so is $R$ and it is convenient to write $R = - e^\beta$ for a real parameter $\beta$. The sign is chosen because it turns out below that $C_-/C_+$ is negative once normalizability is imposed at infinity. In this case \pref{bceps2} becomes
\be \label{bcepsbetaR}
  \frac{\hat \lambda}{\zeta_s} =  \frac{1-R}{1+R}  = -  \coth \frac{\beta}2 \,,
\ee
and our criterion for finding $\hat\lambda(\epsilon)$ is to demand its dependence cancel the explicit $\epsilon$-dependence that is hidden within $R$ (or $\beta$) in \pref{bcepsbetaR}. Differentiating this expression with respect to $\epsilon$ using the $\epsilon$-independence of $\kappa$ and $C_-/C_+$ in \pref{Rexp} to infer $\epsilon\, \exd \beta/\exd \epsilon = -\zeta_s$, leads to the RG equation
\be
 \epsilon \, \frac{\exd}{\exd \epsilon} \left( \frac{\hat \lambda}{\zeta_s} \right) =  \frac{1}{ \sinh^2 ({\beta}/2)} \left( \frac{\epsilon}{2} \,\frac{\exd \beta}{\exd \epsilon} \right) = \frac{\zeta_s}{2} \left[ 1 -\coth^2 \frac{\beta}2  \right]  = \frac{\zeta_s}{2} \left[1 - \left( \frac{\hat \lambda}{\zeta_s} \right)^2 \right] \,.
\ee

\begin{figure}[h]
\begin{center}
 \includegraphics[width=90mm,height=60mm]{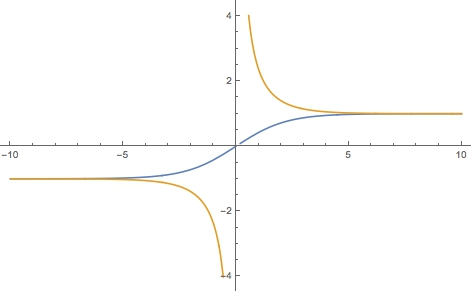} 
\caption{Plot of the RG flow of $\hat\lambda/\zeta_s$ vs $\ln \epsilon/\epsilon_\star$ where $\hat \lambda = (\lambda/2\pi \epsilon) +1$ and with $-\frac 34 < \xi < \frac14$ chosen so that $\zeta_s$ is real. A representative of each of the two RG-invariant classes of flows is shown, and $\epsilon_\star$ is chosen as the place where $\hat \lambda = 0$ or $\hat \lambda \to \infty$, depending on which class of flows is of interest.} \label{fig:RGflow} 
\end{center}
\end{figure}

This RG flow clearly has fixed points at $\hat \lambda = \pm \zeta_s$ and integrates to give
\be \label{lambdaeps}
 \frac{\hat \lambda(\epsilon)}{\zeta_s} =  \frac{ (\hat \lambda_0/\zeta_s) + \tanh\left[ \frac12 \, \zeta_s \ln(\epsilon/\epsilon_0) \right]}{1 + (\hat \lambda_0/\zeta_s) \, \tanh\left[ \frac12 \, \zeta_s \ln(\epsilon/\epsilon_0)\right]} \,.
\ee
This shows how $\hat\lambda$ flows with increasing $\epsilon$ ({\em i.e.} from the UV to the IR) from the fixed point at $-\zeta_s$ when $\epsilon \to 0$ up to $+\zeta_s$ as $\epsilon \to \infty$, passing through the value $\lambda_0$ when $\epsilon = \epsilon_0$. Notice this depends only on the inverse-square coupling through $\zeta_s$, but remains nontrivial even when this coupling vanishes ({\em i.e.} when $2mg = \xi = 0$ and so $\zeta_s = 1$). Of particular later interest is the observation that zero coupling (that is, $\lambda = 0$ and so $\hat \lambda = 1$) is only a fixed point when $2mg = \xi = 0$. Notice also that there are two distinct classes of flows --- as illustrated in Fig.~\ref{fig:RGflow} --- that differ in the RG-invariant criterion of whether $|\hat \lambda/\zeta_s|$ is larger than or smaller than unity. 

Of later interest is the asymptotic form for this running as $\hat\lambda(\epsilon)$ nears the fixed points at $\hat\lambda = \pm\zeta_s$. Using the asymptotic expression for $\tanh z$ for large positive or negative $z$ leads to 
\bea   \label{lamapp}
 \frac{\hat \lambda(\epsilon)}{\zeta_s}   &=& \frac{ (\hat \lambda_0/\zeta_s) + \tanh\left[ \frac12 \, \zeta_s \ln(\epsilon/\epsilon_0) \right]}{1 + (\hat \lambda_0/\zeta_s) \, \tanh\left[ \frac12 \, \zeta_s \ln(\epsilon/\epsilon_0)\right]} \nn\\ 
 &\simeq&   1 +  2\left( \frac{\epsilon_0}{\epsilon} \right)^{\zeta_s} \left( \frac{\hat\lambda_0 -\zeta_s}{\hat\lambda_0 +\zeta_s} \right) + \cO\left[ \left( \frac{\epsilon_0}{\epsilon} \right)^{2\zeta_s} \right]   \quad\quad\; \hbox{(for $\epsilon \gg \epsilon_0$)}  \\ 
 &\simeq&   -1 -  2\left( \frac{\epsilon}{\epsilon_0} \right)^{\zeta_s} \left( \frac{\hat\lambda_0 +\zeta_s}{\hat\lambda_0 -\zeta_s} \right) + \cO\left[ \left( \frac{\epsilon}{\epsilon_0} \right)^{2\zeta_s} \right]     \quad\quad\; \hbox{(for $\epsilon \ll \epsilon_0$)} \,.\nn
\eea
which reveals how the quantity $1-\zeta_s = 1-\sqrt{1 - 4\xi}$ acts as an `anomalous dimension' for $\hat\lambda$.

\subsection{Bound states}

Bound states are found by imposing normalizability of $\psi = C_+ \psi_+ + C_- \psi_-$ at large $r$, which can be written (with arbitrary normalization constant $C$) as
\be
 \psi_\infty(r) =  C \left[  \frac{\Gamma(-\zeta)}{\Gamma\left[\frac12\left(-\frac{w}{\kappa} +1 - \zeta\right)\right]} \; \psi_+(r) +
 \frac{\Gamma(\zeta)}{\Gamma\left[\frac12\left(-\frac{w}{\kappa} +1 + \zeta\right)\right]} \; \psi_-(r) \right]  \,.
\ee
Integer $\zeta$ can be problematic in this expression and so is obtained by a limiting procedure. Clearly this fixes the ratio $C_-/C_+$ to be
\be \label{inftybc}
 \frac{C_-}{C_+} = \frac{\Gamma(\zeta) \Gamma\left[ \frac12 \left( - \frac{w}{\kappa} +1 - \zeta\right)\right]}{\Gamma(-\zeta) \Gamma\left[ \frac12 \left( - \frac{w}{\kappa} +1 + \zeta\right)\right]} = - \frac{\Gamma(1+\zeta) \Gamma\left[ \frac12 \left( - \frac{w}{\kappa} +1 - \zeta\right)\right]}{\Gamma(1-\zeta) \Gamma\left[ \frac12 \left( - \frac{w}{\kappa} +1 + \zeta\right)\right]} \,,
\ee
and so demanding this be consistent with the condition \pref{Rvslamzet} gives the quantization conditions for $\kappa$. 

For all but the $s$-wave we have seen (at least for $-\frac34 \le \xi < \frac54$) that normalizability at $r = 0$ requires $C_- = 0$, so consistency with \pref{inftybc} is not possible at all in the absence of a Coulomb potential ({\em i.e.} when $w = 0$), indicating the absence of a bound state in this case. On the other hand, when $w \ne 0$ consistency requires $\kappa$ must sit at a pole of the denominator, which  ensures
\be \label{kappaquant}
 \kappa = \frac{w}{2N+1+\zeta} \,,
\ee
for $N = 0,1,2,\cdots$. This is also the solution for $s$-wave states if $\hat \lambda = \zeta_s$, since $R = 0$ in this case too.

For the Schr\"odinger Coulomb problem (with no inverse-square potential) we have $g = 0$ and $w = 2mZ\alpha$ while $\zeta = \zeta_c := 2\ell + 1$ where $\ell = 1,2,...$ is the angular momentum quantum number. For all $\ell \ne 0$ states \pref{kappaquant} then returns the usual Schr\"odinger eigenvalues: $E = - \kappa^2/(2m) = - {m (Z \alpha)^2}/({2n^2})$, where the principal quantum number is $n = N+1+\ell \ge \ell + 1$. Eq.~\pref{kappaquant} also captures the Klein-Gordon energy levels once we include also the inverse-square term in the potential. In this case we find $\zeta = 2l + 1$, with the non-integer $l$ now defined by \pref{ldef}, which gives the standard result when inserted into $\omega^2 = - \kappa^2 + m^2$.

\subsubsection*{Perturbing of $s$-wave energies when $\hat\lambda \ne \zeta_s$ }

Consider next the more general $s$-wave case, in the case where $\zeta_s = \sqrt{1 - 4\xi}$ is real. In this case the quantization condition \pref{bcepsbetaR} that determines $\kappa$  has no solutions for RG trajectories satisfying $|\hat \lambda| < \zeta_s$ and for flows with $|\hat \lambda| > \zeta_s$ the solution is found by solving for $\kappa$ in
\be \label{boundcdn}
 \frac{C_-}{C_+} = \left[ \frac{\zeta_s - \hat \lambda}{\zeta_s + \hat \lambda} \right] (2\kappa \epsilon)^{\zeta_s} =  \frac{\Gamma(\zeta_s) \Gamma\left[ \frac12 \left( - \frac{w}{\kappa} +1 - \zeta_s\right)\right]}{\Gamma(-\zeta_s) \Gamma\left[ \frac12 \left( - \frac{w}{\kappa} +1 + \zeta_s\right)\right]}\,.
\ee
As mentioned above, this reduces to the standard Coulomb energy level when the left-hand side vanishes, as it would if either $C_- = 0$ or $\hat \lambda = \zeta_s$. 

An extreme limit occurs when $w = 0$ (so where there is no $1/r$ component to the potential), in which case the solution reduces to the result found in \cite{SchInvSq}: 
\be \label{boundcdnw0}
 \kappa \simeq \frac{1}{\epsilon} \left\{ \frac{\zeta_s + \hat\lambda}{\zeta_s - \hat\lambda} \left[ \frac{\Gamma(\zeta_s) \Gamma\left[ \frac12 \left( 1 - \zeta_s\right)\right]}{\Gamma(-\zeta_s) \Gamma\left[ \frac12 \left( 1 + \zeta_s\right)\right]} \right] \right\}^{1/\zeta_s} \quad \hbox{when $w \simeq 0$}\,.
\ee
Physically, because $w = 0$ and $\xi < \frac14$ this bound state is dominantly supported by the delta-function potential furnished by the contact interaction whose strength is governed by $\lambda$. 

When $w \ne 0$ a useful formula for how energy levels are perturbed from their Coulomb (or Klein-Gordon) limit when $\hat \lambda- \zeta_s$ is not too large is found by approximating the gamma-function near its pole by $\Gamma(z-N)  \simeq \frac{(-)^N}{N! \, z} \Bigl[1 + \cO(z) \Bigr] $, where $z$ is near zero. Using this when $\kappa$ is near a zero of $C_-$ we find \pref{boundcdn} takes the approximate form
\bea \label{singlepole}
    \left( - \frac{w}{2\kappa} +\eta \right) \frac{1}{(2\kappa\epsilon)^{\zeta_s}} &\simeq& \frac{\Gamma(\zeta_s+N+1) }{N! \Gamma(\zeta_s) \Gamma(\zeta_s+1)} \left(\frac{\zeta_s - \hat \lambda}{\zeta_s + \hat \lambda} \right)  \nn\\
    &\simeq& \frac{\Gamma(\zeta_s+n) }{(n-1)! \Gamma(\zeta_s) \Gamma(\zeta_s+1)}  \left( \frac{\epsilon_0}{\epsilon} \right)^{\zeta_s} \frac{\zeta_s- \hat\lambda_0}{\zeta_s+\hat\lambda_0} \,,
\eea
for $\eta = N+1+l$ and $-l = \frac12(1-\zeta_s)$ as above, with $N = n-1 = 0,1,2,\cdots$ corresponding to the principal quantum number $n$ of the Coulomb limit, as above. The second line assumes $\hat\lambda(\epsilon)$ is specified by giving its value $\hat\lambda_0 = \hat\lambda(\epsilon_0)$ at some microscopic scale $\epsilon_0$, and uses the asymptotic expression \pref{lamapp}. Notice the cancellation of the explicit $\epsilon$-dependence in this formula.

The solution perturbatively close to the zeroth order solution of the Coulomb/inverse-square problem is $\kappa = \overline \kappa + \delta \kappa = ({w}/{2\eta}) + \delta \kappa$ with $\delta \kappa$ given by
\be \label{deltakappa}
  \frac{\delta \kappa}{\overline \kappa} \simeq  \frac{ \left( 2\overline \kappa\epsilon_0 \right)^{\zeta_s}}{\eta}  \left( \frac{\zeta_s- \hat\lambda_0}{\zeta_s+\hat\lambda_0} \right)  \frac{\Gamma(\zeta_s+n) }{(n-1)! \Gamma(\zeta_s) \Gamma(\zeta_s+1)}  \,.
 \ee

Of course, the mere existence of a solution for $\kappa$ does not suffice to ensure the presence of a physical bound state. In order to be trusted the bound state must be much larger than the UV scale that characterizes the structure of the source, and which provides a lower limit to the length scales for which an analysis purely within the point-particle EFT can be valid. For the Coulomb-like solutions  the size of the bound state is given as usual by the `Bohr radius', or $r \sim w^{-1}$ where $w = 2ms (= 2mZ \alpha)$. Believability of the bound state requires $\kappa \epsilon \ll 1$ where $\epsilon$ is a UV scale.

For bound states where the contact interaction plays an important role demanding the bound state be much larger than UV scales imposes a condition on $\lambda$, and this is how we see why the delta-function potential must be attractive and sufficiently strong. To see how this works we must identify the scale of the bound state determined by \pref{boundcdnw0}, and this is most simply identified by exploiting the $\epsilon$-independence of equations like \pref{boundcdnw0} to express the result in terms of an RG-invariant scale. Since $|\hat\lambda| > \zeta_s$ it is natural to choose this RG-invariant scale to be the scale $\epsilon_\star$ where $|\hat \lambda (\epsilon_\star)| = \infty$, leading to
\be \label{boundcdnw0RG}
 \kappa \simeq \frac{1}{\epsilon_\star} \left| \frac{\Gamma(\zeta_s) \Gamma\left[ \frac12 \left( 1 - \zeta_s\right)\right]}{\Gamma(-\zeta_s) \Gamma\left[ \frac12 \left( 1 + \zeta_s\right)\right]}  \right|^{1/\zeta_s} \quad \hbox{when $w \simeq 0$}\,.
\ee 
For generic $\zeta_s$ this shows the bound state is of order $\epsilon_\star$ in size. To be trusted for any UV scale $\epsilon$ on the RG flow we must ask $\hat\lambda(\epsilon)$ to be such that $\epsilon_\star \gg \epsilon$. Taking $\hat\lambda_0 \to \infty$ in the RG flow \pref{lambdaeps} implies 
\be
 \frac{\hat \lambda(\epsilon)}{\zeta_s} =  \coth\left[ \frac12 \, \zeta_s \ln(\epsilon/\epsilon_\star)\right] \,,
\ee
and so demanding $\epsilon \ll \epsilon_\star$ implies $\hat\lambda(\epsilon) \simeq - \zeta_s(1 + \delta)$ with $0 < \delta \ll 1$, and it is only for such couplings in the UV that a macroscopic bound state of the form \pref{boundcdnw0} can be trusted.

\subsection{Scattering}

Scattering calculations go through in a very similar way, and for later purposes we collect results here for the scattering amplitude, restricted to the case $-\frac 34 \le \xi \le \frac14$ for which $\zeta_s$ is real. Our treatment here follows that of \cite{SchInvSq} fairly closely.

The scattering result also shows how renormalization makes the contribution to scattering of the contact interaction, $\lambda$, depend only on RG-invariant scales like $\epsilon_\star$, rather than being set directly by the microscopic scale $\epsilon$ where $\lambda(\epsilon)$ is matched to the UV completion of the source. This can make scattering effects surprisingly large in those circumstances where $\epsilon_\star \gg \epsilon$. 

As before the starting point is the radial solution in the form $\psi = C_+ \psi_+ + C_-\psi_-$, with $C_-/C_+$ set by the boundary condition as $r \to 0$. For the range of $\xi$ considered here this boundary condition ensures $C_- = 0$ for all $\ell \ne 0$, while the $s$-wave state satisfies \pref{Rvslamzet}, which states
\be \label{Rvslamzetz}
 \frac{C_-}{C_+} = R(\hat\lambda/\zeta_s) (2ik \epsilon)^{\zeta_s} = \left[ \frac{\zeta_s - \hat \lambda}{\zeta_s + \hat \lambda} \right] (2ik \epsilon)^{\zeta_s} \,,
\ee
which also writes $\kappa = ik$, as appropriate for a state with $E = k^2/2m > 0$. Unlike for bound states the ratio $C_-/C_+$ is not independently set by normalizability at large $r$. Notice  it is again the {\em difference} between $\hat \lambda$ and its IR fixed point value that drives $C_-/C_+$ away from what would be found in the absence of a contact interaction with the source ({\em i.e.} drives it away from $C_- = 0$).

Evaluating asymptotically close to the IR fixed point at $\hat \lambda = \zeta_s$ using \pref{lamapp} and inserting into \pref{Rvslamzetz} we see the expected cancellation of powers of $\epsilon$ leaving
\be \label{ratioapprox}
  \frac{C_-}{C_+}  \simeq \left(2ik\epsilon_0 \right)^{\zeta_s}   \frac{\zeta_s - \hat\lambda_0}{\zeta_s + \hat\lambda_0}  = - y \, (2i k \epsilon_\star)^{\zeta_s} \,,
\ee
where the last equality uses the RG-invariant scale $\epsilon_\star$, defined by $\hat\lambda(\epsilon_\star) = \infty$ (if $|\hat\lambda_0| > \zeta_s$) or $\hat \lambda(\epsilon_\star) = 0$ (if $|\hat \lambda_0| < \zeta_s$). Here $y = \hbox{sign}[|\hat \lambda|-\zeta_s]$ is the RG-invariant sign that determines which of these definitions of $\epsilon_\star$ is to be used.

To match $C_-/C_+$  to the scattering amplitude we write the large-$r$ behaviour of our wavefunction as
\be
\label{LL1}
\psi \to A_\ell \, \frac{e^{i(kr - \ell\pi/2)}} r + B_\ell \, \frac{e^{-i(kr - \ell\pi/2)}} r \,,
\ee
and define the phase shift by \cite{LL} $e^{2i\delta_\ell} = - {A_\ell}/ {B_\ell} $. Taking the large-$r$ limit of the confluent hypergeometric function leads to 
\bea
\label{andSo}
\psi_\pm &\propto& e^{- ikr} \, \frac {\Gamma(1\pm \zeta)}{\Gamma\left[\frac 12\left(-iw/k + 1 \pm \zeta\right)\right]}\, e^{\frac \pi 2 \left[ i\left(1 \pm \zeta\right)-w/k \right] } (2ikr)^{-1-iw/2k } \notag \\
&&\qquad\qquad\qquad {}+ e^{ikr}\, \frac {\Gamma(1 \pm \zeta)}{\Gamma\left[\frac 12\left(iw/k + 1 \pm \zeta\right)\right]} (2ikr)^{-1+iw/2k},
\eea
which permits reading off the phase shift. 

For large $r$ we drop oscillating factors like $(2kr)^{\pm w/2ik} = e^{\mp i (w/2k) \ln(2kr)}$ that are subdominant to the exponentials $e^{\pm ikr}$, leading for $\ell \ne 0$ (and for $\ell = 0$ when $\hat \lambda = \zeta_s$) to the phase shift 
\bea
\label{phasene0}
e^{2i\delta_\ell} &=&  \frac {\Gamma\left[\frac 12\left(-\frac {iw} {k} + 1 + \zeta\right)\right] }{\Gamma\left[ \frac 12\left(\frac {iw} {k} + 1 + \zeta\right)\right] }   \; e^{i\pi (\ell-l)} \,,
\eea
which uses $\zeta = 2l + 1$. Notice that in the absence of an inverse-square potential ($\xi = 0$ and so $l = \ell$) this expression reduces to the usual one for Rutherford scattering \cite{LL}
\be
\label{close}
e^{2i\delta_\ell}  =  \frac {\Gamma(\ell + 1 - iw/2k)}{\Gamma(\ell + 1 + iw/2k)} \qquad \hbox{(Rutherford limit)} \,.
\ee
On the other hand, for $s$-wave scattering in general $C_-/C_+$ is given by \pref{ratioapprox}, leading to
\bea
\label{phase0}
e^{2i\delta_0} &=& \frac {\Gamma(1 + \zeta_s)/\Gamma\left(\frac 12\left(\frac{i w} {k} + 1 + \zeta_s\right)\right) -y (2ik\epsilon_\star)^{\zeta_s}\Gamma(1 - \zeta_s)/\Gamma\left(\frac 12\left(\frac {iw} {k} + 1 - \zeta_s\right)\right)}{\Gamma(1 + \zeta_s)/\Gamma\left(\frac 12\left(-\frac {iw} {k} + 1 + \zeta_s\right)\right) -y (-2ik\epsilon_\star)^{\zeta_s}\Gamma(1 - \zeta_s)/\Gamma\left(\frac 12\left(-\frac {iw} {k} + 1 - \zeta_s\right)\right)} \; e^{(1 - \zeta_s)i\pi /2} \,.\nn\\
&&
\eea

Of later interest is the case where the Coulomb contribution is turned off, and so for which $w = 0$. In this case --- as shown in more detail in  \cite{SchInvSq} --- the scattering phase shift simplifies to become  
\be
\label{phaseA}
e^{2i\delta_0} = \left[ \frac{1 - \cA \,e^{i\pi \zeta_s/2}}{1 -\cA \,e^{-i \pi \zeta_s/2} } \right] \; e^{(1 - \zeta_s)i\pi /2} \qquad \hbox{($w = 0$ limit)} \,,
\ee
where
\be
  \cA := y \left( \frac{k \epsilon_\star}{2} \right)^{\zeta_s} \left[ \frac{\Gamma\left(1 - \frac12 \, \zeta_s \right)}{ \Gamma\left(1 + \frac12 \, \zeta_s \right) } \right] \,.
\ee

A final limit is the case of scattering from a delta-function, obtained by turning off the inverse-square potential and taking $\xi = 0$ and $\zeta_s = 1$. In this limit we have
\be
\label{phasedelta}
e^{2i\delta_0} =  \frac{1 -i \cA_\delta }{1 +i\cA_\delta  }  \qquad \hbox{($\delta$-function scattering)} \,,
\ee
with
\be
  \cA_\delta := y \left( \frac{k \epsilon_\star}{2} \right)  \frac{\Gamma\left(\frac12 \right)}{ \Gamma\left(\frac32 \right) } =  y k \epsilon_\star \,.
\ee
This agrees with standard calculations \cite{Jackiw} and in particular gives $\tan \delta_0 = - \cA_\delta$. At low energies the scattering length, $a_s$, is given by $k \cot \delta_0  \simeq - 1 / a_s + \cO(k^2)$ (so that the low-energy cross section is $\sigma = 4\pi a_s^2$). When the $\delta$-function dominates in the scattering we therefore find $a_s$ directly fixes the RG-invariant scale through the relation
\be \label{asvseps}
  a_s = y \epsilon_\star \,.
\ee

\section{Applications}
\label{sec:Apps}

We now turn to several practical applications to the developments of the previous section. These include the reproduction and clarification of some well-known results (such as the Deser formula relating the energy-level shift and scattering length of pion-nucleon interactions in pionic hydrogen states); a brief recap of the argument of \cite{SchInvSq} as to why classical renormalization provides a simple and intuitive low-energy description for how scattering from small objects like magnetic monopoles can catalyze reactions; a treatment of mixing induced by contact interactions; and a discussion of how the interplay of relativistic effects with the classical renormalization of contact interactions can amplify the size of contact interactions within mesonic atoms. Although (as shown in a companion paper \cite{Dirac}) some of these features also carry over to a Dirac-equation treatment including spin, this is not so for the spectacular energy level shift of order $\epsilon_\star/m$.

\subsection{Pionic atoms and the Deser formula}

For our first application we consider mesonic (pionic and kaonic) atoms, in which a relatively long-lived and negatively charged meson orbits a nucleus (or proton, in the simplest case). Such `atoms' are of interest because the mesons live long enough to be captured by the nucleus once a beam is brought to impinge on a target material. Once captured, the meson cascades down to the ground state and detection of the X-rays emitted in this process allows the measurement of the bound-state energies. 

The binding is electromagnetic because the mesonic Bohr radius is much larger than the range of nuclear forces (that are set by the pion Compton wavelength) and because of this $v^2/c^2 \sim \alpha$ is small enough to be well within the non-relativistic regime. Additionally, since the meson mass, $m$, is at least 300 times larger than the electron mass its orbital radius is at least 300 times smaller, bringing the mesonic orbit well inside the various electronic ones. In this case the influence of meson-nuclear strong forces can be modelled by a contact interaction in an effective theory that does not resolve the nuclear size, making the formalism of this paper appropriate. Measurements of the energy-level shift induced by the meson-nucleon strong interaction probe the detailed nature of meson-nucleon interactions \cite{pionicatoms, kaonicatoms}.

In this section we use the previously presented formalism to derive the Deser formula \cite{Deser} relating the strong-interaction shift to the mesonic bound state energy to the meson-nucleus scattering length. This formula usually is derived using a model for the nuclear potential acting over short distances, in which the need to go beyond Born approximation is often emphasized. Our presentation here shows how the discussion naturally fits within the framework of a point-particle EFT and how the need for contributions beyond Born approximation are captured in a controlled way by the RG evolution of the nuclear contact interaction.

The starting point is the Schr\"odinger action \pref{Sbulk} coupled to a contact interaction, \pref{Ssource}, meant to represent the short-range strong meson-nucleon interactions. We parameterize this interaction here in terms of the coupling $h$, as above, though a more systematic exploration of the kinds of contact interactions possible might also be warranted.\footnote{Much thought has been put into the meson-nucleon effective interaction within chiral perturbation theory, in which the dominant term is momentum-dependent though smaller momentum-independent Yukawa-style interactions are also possible \cite{ProtonRadiusStandards}. For the purposes of illustration we restrict ourselves here to a simple Yukawa interaction, though a more sophisticated and systematic treatment is clearly possible.}

In this case the results of \S\ref{sec:NRmixedCISP} can be taken over in whole cloth, and neglecting very small relativistic effects (more about which below) we can take the Coulomb potential to have strength $s = Z \alpha$ and the inverse-square potential to vanish: $g = 0$. The quantization condition that sets the binding energy of the hydrogen-like mesonic state is then given by \pref{boundcdn}, in which we use $w = 2ms = 2m Z \alpha$ and $\lambda = 2m h$ while the condition $\xi = 2m g = 0$ ensures $\zeta_s = 1$. Because we set $g = 0$ (and so have no inverse-square potential) it is RG-invariant to choose $h = 0$, although in this case we do not do so because its value captures a physical effect: the strength of the short-range meson-nucleon force. 

In the regime of interest the bound-state condition is solved by a relatively small change from the Schr\"odinger Coulomb solution as in \pref{deltakappa}, leading to 
\be \label{deltakappamesonic}
  \frac{\delta \kappa}{\overline \kappa} \simeq    2\,\overline \kappa\,\epsilon_0    \left( \frac{1- \hat\lambda_0}{1+\hat\lambda_0} \right)  = - 2\, y   \left( \frac{\epsilon_\star}{na_\ssB} \right) \quad \hbox{(if $\ell = 0$)}\,.
\ee
Here $n$ is the principal quantum number and $a_\ssB = (mZ \alpha)^{-1}$ is the mesonic Bohr radius, while $h_0 \simeq \epsilon_0^2$ is a typical nuclear scale when specified at nuclear distances, $\epsilon_0 \simeq 1$ fm. This ensures $\hat\lambda_0 \simeq \cO(1)$ and so also that $\epsilon_\star$ --- defined as the scale where $\hat \lambda$ diverges (if $|\hat \lambda_0| > 1$) or where $\hat \lambda = 0$ (if $|\hat \lambda_0| < 1$) is also a typical nuclear size $\epsilon_\star  \simeq \epsilon_0$. (As in previous sections $y = \hbox{sign}[|\hat\lambda_0| - 1]$ is the RG-invariant sign that distinguishes the two types of RG flow.) 

This leads to the following shift in the mesonic bound state energy,
\be \label{dEpion}
 \delta E_n = - \delta \left( \frac{\kappa^2}{2m} \right) = - \frac{\bar \kappa \, \delta \kappa}{m} =  2\,y    \left( \frac{\epsilon_\star}{m n^3a_\ssB^3} \right) \quad \hbox{($s$-wave only)}  \,.
\ee
As usual the size of the influence of the contact interaction on physical quantities is set by the RG-invariant scale $\epsilon_\star$ found from its coupling $\lambda$. In the present instance this is generically similar in size to the nuclear scale, $\epsilon_0$, at which matching to the UV completion describing the nucleus occurs. 

But in the end, both $\lambda_0$ and $\epsilon_\star$ are just parameters, and a real prediction comes only once they are traded for another observable. One such observable is the scattering length, $a_s$, of mesons from nucleons, which if governed at low-energies by the same contact interaction is given by \pref{asvseps}, or $y \epsilon_\star = a_s$. Using this in \pref{dEpion} leads to the following relationship between the fractional strong-interaction shift in the $s$-wave energy levels of mesonic atoms to the low-energy elastic scattering length for mesons scattering from the same nucleus:
\be \label{dEdeser}
 \frac{\delta E_n}{|E_n|} \simeq 2 \left( \frac{\delta \kappa}{\bar \kappa} \right) = \frac{4a_s}{na_\ssB}  \quad \hbox{($s$-wave only)}  \,.
\ee
For the ground state $n =1$ this reproduces the Deser formula \cite{Deser} for mesonic atoms. As is usual for an EFT analysis, corrections to this expression should arise from higher-dimension interactions localized at the source, and because of their higher dimension would be expected to be suppressed by further powers of $\epsilon_\star/a_\ssB$. 

We see that for mesonic atoms it is well-known that energy shifts can receive contributions linear in a microscopic UV scale. 

\subsection{RG scales and reaction catalysis}

For completeness we briefly reiterate here a point made in \cite{SchInvSq} concerning reaction catalysis.

In some problems the scattering of interest between a particle and a point source is dominated by the $\delta$-function contact interaction $h \, \delta^3(\bfx)$, rather than the longer-range Coulomb or inverse-square potentials. When this is true, \pref{phasedelta} and \pref{asvseps} show that the low-energy cross section is  $\sigma \simeq 4\pi a_s^2$ where the scattering length is of order the RG-invariant scale $\epsilon_\star$ set by the classical running of $h$. The value of $\epsilon_\star$ is in turn predictable from the RG evolution in terms of any initial condition $h(\epsilon_0) = h_0$ that might fix $h$ at a UV scale $\epsilon_0$, perhaps where the low-energy point-particle EFT is matched to whatever UV completion describes the source's internal structure. 

Now comes the main point. Although it is often the case that $\epsilon_\star$ is of order the geometrical size $\epsilon_0$ suggested by such a matching (such as was found for mesonic atoms in the previous example), it can also happen that $\epsilon_\star $ differs considerably, with $\epsilon_\star \gg \epsilon_0$ when $\hat \lambda_0$ is very close to the UV fixed point (at $\hat \lambda = - \zeta_s$) or with $\epsilon_\star \ll \epsilon_0$ when $\hat \lambda_0$ is close to the IR fixed point (at $\hat \lambda = + \zeta_s$). In particular, if the UV theory happens to match to the effective theory at $\epsilon = \epsilon_0$ with $h \simeq -(\pi \epsilon/m)(1+\zeta_s)$ then because this ensures $\hat \lambda \simeq -\zeta_s$ it also guarantees that $\epsilon_\star \gg \epsilon_0$. In such a case the low-energy scattering cross section can be much larger than the geometrical one suggested by the UV scale $\epsilon_0$. 

As discussed in \cite{SchInvSq} a concrete case where we believe these observations to apply is to $s$-wave scattering of charged particles from magnetic monopoles \cite{monopolerev}. The radial equation studied here applies to the non-relativistic limit (and --- see below --- to the relativistic case for spinless particles), though in general such scattering also involves an inverse-square potential because the magnetic monopole alters the particle angular momentum. In particular, for spinless particles the angular part of the problem alters the angular-momentum quantum number away from a non-negative integer to $\ell = \mu, \mu+1, \cdots$ where $\mu = eg/4\pi = \hat n/2$ with $g$ the monopole's magnetic charge and $e$ the electric charge of the scattering particle (and the relation to an integer $\hat n$ is as required by the Dirac quantization condition). In terms of these quantities the dimensionless coefficient, $-v$, of the inverse-square potential is $l(l+1) = \ell(\ell+1) - \mu^2$ and so $\xi = \mu^2$. 

These expressions show that when $\mu = \frac12$ (say) there is no value of angular quantum numbers for which $l(l+1)$ vanishes, so the inverse-square coupling always plays a role. But the same exercise shows that for spin-$\frac12$ particles there is an $s$-wave combination for which the spin combines with $\mu = \frac12$ to allow $v = 0$, in which case the scattering is purely governed by the $\delta$-function component. As we see above (and is argued in \cite{SchInvSq}), this opens the possibility for cross sections being much larger than geometric in size provided the matching in the UV provides a coupling $h_0$ in the right range. This leaves open (see, however, \cite{Dirac}) why the standard arguments associated with monopole catalysis of baryon-number violation \cite{monopolecatal} provide the microscopic UV boundary conditions required to enhance scattering cross sections, thereby allowing classical RG evolution to provide a simple explanation for the unexpectedly large size of these cross sections.

\subsection{Mixing through contact interactions}

Since we have seen that renormalization can cause contact interactions to cause surprisingly large effects, one might ask whether this renormalization floats all boats and amplifies {\em all} possible contact interactions. This section explores this issue by considering the RG evolution of contact interactions for two species of particles and shows why for some contact interactions zero coupling remains a fixed point even in the presence of an inverse-square potential. The interactions that are not amplified do not share the same selection rules as does the inverse-square potential itself, and this is what decides which interactions become enhanced. 

To explore this further imagine extending the Schr\"odinger field to a 2-component Pauli field, 
\be
 \Psi = \left( \begin{array}{ccc} \psi_1  \\ \psi_2   \end{array} \right) \,,
\ee
on which internal $SU(2)$ `flavour' rotations are represented by the usual Pauli matrices. We take the bulk description to be $SU(2)$-invariant but imagine this symmetry to be broken by the source action, which is taken to be
\be
 S_b = - \int \exd \tau \Bigl[ h_0 \, \Psi^\dagger \Psi + h_3 \, \Psi^\dagger \sigma_3 \Psi \Bigr]
 = - \int \exd \tau \Bigl[ (h_0 + h_3) \, \psi_1^* \psi_1 + (h_0 - h_3) \, \psi_2^* \psi_2 \Bigr] \,.
\ee
As usual we define $\lambda_0 = 2m h_0$ and $\lambda_3 = 2m h_3$.

Repeating the argument given above for each of $\psi_1$ and $\psi_2$ returns precisely the same boundary condition as before:
\be
 \left. 4 \pi \epsilon^2 \, \frac{\partial \psi_1}{\partial r} \right|_{r = \epsilon} = (\lambda_0 + \lambda_3) \, \psi_1(\epsilon) 
 \quad \hbox{and} \quad
 \left. 4 \pi \epsilon^2 \, \frac{\partial \psi_2}{\partial r} \right|_{r = \epsilon} = (\lambda_0 - \lambda_3) \, \psi_2(\epsilon) \,,
\ee
and because $\partial_r \ln \psi (\epsilon)$ is a function of $\zeta_s$ and $\kappa \epsilon$ that depends only on the bulk field equations and how their radial solutions approach the origin, the RG equation found by differentiating the above with respect to $\epsilon$ is also the same as found in earlier sections:
\be
 \epsilon \, \frac{\exd}{\exd \epsilon} \left( \frac{\hat \lambda_\pm}{\zeta_s} \right) = \frac{\zeta_s}{2} \left[ 1 - \left( \frac{\hat \lambda_\pm}{\zeta_s} \right)^2 \right] \,,
\ee
where
\be
 \hat \lambda_\pm = \frac{\lambda_0 \pm \lambda_3}{2 \pi \epsilon} + 1 \,.
\ee
Notice that when $\zeta_s = 1$ the fixed point for these flows occurs at $\lambda_0 = \mp \lambda_3$, which is the case where $S_b$ projects out either $\psi_1$ or $\psi_2$. 

Suppose we define
\be
 \hat \lambda_0 := \frac{\lambda_0}{2\pi \epsilon} \quad \hbox{and} \quad
 \hat \lambda_3 := \frac{\lambda_3}{2\pi \epsilon} \,,
\ee
so that $\hat \lambda_\pm = \hat \lambda_0 \pm \hat \lambda_3 + 1$. Then the evolution for these two new variables is given by
\be
 \epsilon \, \frac{\exd}{\exd \epsilon} \left( \frac{\hat \lambda_0}{\zeta_s}  \right) = \frac{\zeta_s}{2} \left[ 1 -  \frac{(\hat \lambda_0 + 1)^2 + \hat \lambda_3^2}{\zeta_s^2} \right] \,,
\ee
and 
\be
 \epsilon \, \frac{\exd}{\exd \epsilon} \left( \frac{ \hat\lambda_3}{\zeta_s}  \right) = -\left[ \frac{ \hat \lambda_3 (\hat \lambda_0 + 1) }{\zeta_s} \right]  \,.
\ee
Notice that $\hat\lambda_3 = 0$ is a fixed point of this last equation, indicating that it is RG-invariant for this coupling to vanish, even if $\hat \lambda_0 \ne 0$ and $\zeta_s \ne 1$. In particular, when $\hat\lambda_0 = 0$ the RG equation for $\hat\lambda_3$ integrates to give $\hat \lambda_3(\epsilon) = \hat \lambda_3(\epsilon_0) (\epsilon_0/\epsilon)$, which states that $\lambda_3 = 2\pi \epsilon \, \hat\lambda_3$ is $\epsilon$-independent.

The general solutions to the RG equation are given by the same flows as found earlier, for $\hat\lambda_\pm$:
\be \label{lambdapm}
 \frac{\hat \lambda_\pm(\epsilon)}{\zeta_s} = \frac{ [\hat \lambda_\pm(\epsilon_0)/\zeta_s] + \tanh\left[ \frac12 \, \zeta_s \ln(\epsilon/\epsilon_0) \right]}{1 + [\hat \lambda_\pm(\epsilon_0)/\zeta_s] \, \tanh\left[ \frac12 \, \zeta_s \ln(\epsilon/\epsilon_0)\right]} =   \coth\left[ \frac{\zeta_s}2 \, \ln \left( \frac{\epsilon}{\epsilon_{\star\pm}} \right) \right] \,,
\ee
where the second equality specializes the reference point to $\epsilon_\star$, for which $\lim_{\epsilon_0 \to \epsilon_{\star\pm}} \hat\lambda_\pm(\epsilon_0) = \infty$. Consequently
\be \label{lambda3run}
 \frac{1+\hat \lambda_0(\epsilon)}{\zeta_s} = \frac{\hat \lambda_+(\epsilon)+\hat \lambda_-(\epsilon)}{2\zeta_s} = \frac12 \left\{  \coth\left[ \frac{\zeta_s}2 \, \ln \left( \frac{\epsilon}{\epsilon_{\star+}} \right) \right] + \coth\left[ \frac{\zeta_s}2 \, \ln \left( \frac{\epsilon}{\epsilon_{\star-}} \right) \right] \right\}\,,
\ee
and
\be \label{lambda0run}
 \frac{\hat \lambda_3(\epsilon)}{\zeta_s} = \frac{\hat \lambda_+(\epsilon) - \hat\lambda_-(\epsilon)}{2\zeta_s} = \frac12 \left\{  \coth\left[ \frac{\zeta_s}2 \, \ln \left( \frac{\epsilon}{\epsilon_{\star+}} \right) \right] - \coth\left[ \frac{\zeta_s}2 \, \ln \left( \frac{\epsilon}{\epsilon_{\star-}} \right) \right] \right\}\,,
\ee

Since $\coth x \to 1$ for $x \to \infty$ these enjoy the IR fixed points
\be
 \lim_{\epsilon \to \infty} \hat \lambda_0 +1 = \zeta_s 
 \quad \hbox{and} \quad
 \lim_{\epsilon \to \infty} \hat \lambda_3 = 0 \,,
\ee
showing that it is only $\hat\lambda_0$ that is driven away from zero when $\zeta_s \ne 1$.  For $\epsilon \gg \epsilon_{\star\pm}$ we use $\coth x \simeq 1 + 2 e^{-2x} + \cdots$ to infer the following approach to the IR fixed points:
\be \label{lambda3runap}
 \frac{1+\hat \lambda_0(\epsilon)}{\zeta_s} \simeq  1 + \left( \frac{\epsilon_{\star+}}{\epsilon} \right)^{\zeta_s} + \left( \frac{\epsilon_{\star-}}{\epsilon} \right)^{\zeta_s} + \cdots \,,
\ee
and
\be \label{lambda0runap}
 \frac{\hat \lambda_3(\epsilon)}{\zeta_s} \simeq \left( \frac{\epsilon_{\star+}}{\epsilon} \right)^{\zeta_s} - \left( \frac{\epsilon_{\star-}}{\epsilon} \right)^{\zeta_s} + \cdots  \,.
\ee

The upshot is this: because zero coupling remains a fixed point for $\lambda_3$ even in the presence of an inverse-square potential, it need not be driven to run as dramatically as does the coupling $\lambda_0$. They differ in this way because $\lambda_0$ shares the selection rules of the inverse-square potential while $\lambda_3$ does not. As the above arguments show, rather than implying a complete absence of evolution the RG effects are instead suppressed to enter at higher order in $Z\alpha$. 

\subsection{Klein-Gordon Coulomb problem}

We now argue why the non-relativistic Schr\"odinger analysis given above also carries over directly to a relativistic spinless particle moving in the presence of a Coulomb potential. (We discuss the case of spin-$\frac12$ relativistic particles in \cite{Dirac}.) In particular, the interaction of relativistic particles from point sources turns out to provide a practical example of competing Coulomb and inverse-square potentials, with the Coulomb potential arising with coefficient of order $Z \alpha$ and the inverse-square potential arising due to relativistic effects with a coefficient of order $(Z\alpha)^2$. 

The significance of having both Coulomb and inverse-square potentials in this case is that this ensures that $\xi \simeq (Z\alpha)^2 \ne 0$ and so $\zeta_s \ne 1$. As a result zero-coupling, $h = 0$, is {\em not} a fixed point of the RG evolution of the contact interaction, with the consequence that such a contact interaction {\em must} be nonzero for all scales except perhaps for a specific scale, $\epsilon_0$, at which point $\hat \lambda(\epsilon_0) = 1$. This makes the presence of a contact interaction mandatory, rather than optional, in relativistic Coulomb problems.

\subsubsection*{Relativistic field equation}

The Klein-Gordon equation for a Coulomb potential is given by
\be
(  D_\mu D^\mu - m^2 ) \phi = 0 
\ee
where $D_\mu = \partial_\mu +ie A_\mu$ (for a charge $Q=-1$ particle). Assuming the only nonzero gauge potential to be $eA_0(\bfx) = -Z\alpha/r$ and choosing a stationary state, $\phi(\bfx, t) = \varphi(\bfx) \, e^{-i\omega t}$, this becomes
\be
 0 = \Bigl[- (\partial_t + ieA_0)^2  + \nabla^2  - m^2\Bigr] \phi 
 = \Bigl[ \nabla^2 -2\omega eA_0 + (eA_0)^2  - \kappa^2 \Bigr] \varphi \,,
\ee
where $\omega^2 - m^2 = - \kappa^2$ and for bound state solutions (for which $\omega < m$) we take $\kappa$ to be real. This has the same form as \pref{SchE} --- {\em i.e.} $\nabla^2 \phi - U \, \phi = \kappa^2 \phi$ --- with potential $U(\bfx)$ given by
\be
 U(r) = 2\omega eA_0 - (eA_0)^2 = - \frac{2\omega Z\alpha}{r} - \frac{(Z\alpha)^2}{r^2} \,,
\ee
and so the parameters $v$ and $w$ are
\be
 w = 2 \omega Z \alpha \quad \hbox{and} \quad
 v = (Z\alpha)^2 - \ell(\ell+1) \,,
\ee
which gives $\xi = (Z\alpha)^2$ and
\be
 \zeta = \sqrt{(2\ell + 1)^2 - 4(Z\alpha)^2} \,.
\ee
We see the radial part of the KG equation has the form considered earlier, specialized to these choices for $v$ and $w$. In particular, for $s$-wave states we have 
\be
  \zeta_s = \sqrt{1 - 4(Z\alpha)^2} \simeq 1 - 2 (Z \alpha)^2 \,.
\ee 

\subsubsection*{Boundary conditions}

Because a canonically normalized Klein-Gordon field has dimensions of mass, a contact interaction like $\cL_b = -   h_\KG\, \phi^*\phi \, \delta^3(\bfr)$ has coupling $h_\KG$ with dimension length. Following the steps of \cite{SchInvSq} and integrating over a small Gaussian pillbox to obtain the boundary condition implied for this interaction gives 
\be
  4\pi r^2 \left( \frac{\partial \phi}{\partial r} \right)_{r = \epsilon} = \lambda \phi \,,
\ee
with $\lambda = h_\KG$. This is consistent with the result $\lambda = 2mh$ found for the Schr\"odinger case because canonical normalization of the Schr\"odinger field, $\psi$, requires it to be related to $\phi$ by $\psi = \sqrt{2m}\; \phi$, and so $\cL_b = - h \, \psi^* \psi \, \delta^3(\bfr)$ with $h_\KG = 2mh$. 

The renormalization described earlier goes through as before for $\lambda$, and (as also noted earlier) because $\zeta_s  < 1$ it is inconsistent to choose $h=0$ for all scales. Should we happen to know $h_0 = 0$ at some UV scale $\epsilon_0$ then the flow towards the IR fixed point is given by
\be \label{RGlambda}
 h_\KG = 2mh = \lambda \simeq 2 \pi \epsilon \left\{ - 1 + \zeta_s   \left[  1 +  2 \left( \frac{\epsilon_0}{\epsilon} \right)^{\zeta_s} + \cdots \right] \right\} 
 \quad \hbox{(for $\epsilon \gg \epsilon_0$)} \,.
\ee

\subsection*{Energy shifts in mesonic atoms} 

Any departure of $\hat\lambda$ from $\zeta_s$ implies a deviation from the standard energy-eigenvalue predictions, at least for $s$-wave states, and the surprise is that this is also true in particular if $h_0=0$ at some scale.

To second order in $\epsilon$, the mode functions \eqref{psipm} specialized to the Klein-Gordon Coulomb problem take the form
\begin{align}
 \begin{aligned}
  \psi_+ &\simeq (2\kappa\epsilon)^{\frac12(-1+\zeta_s)} \left( 1 - Z\alpha m \epsilon + \frac{2n^2+1}{6n^2} (Z\alpha m \epsilon)^2 + \mathcal{O}((Z\alpha m \epsilon)^3), \right)\\
  \psi_- &\simeq (2\kappa\epsilon)^{\frac12(-1-\zeta_s)} \left(1 - \frac{m \epsilon}{Z\alpha}  + (m\epsilon)^2 - \frac{2n^2+1}{6n^2} Z\alpha (m\epsilon)^3 + \mathcal{O}((Z\alpha)^2 (m \epsilon)^4)\right), \\
 \end{aligned}
\end{align}
using $1-\zeta_s \simeq 2(Z\alpha)^2$ and $\kappa = \sqrt{(m-\omega)(m+\omega)} \simeq Z\alpha m/n$, where $n$ is the principal quantum number. Combining this with the higher order pole approximation \eqref{deltakappadoublepole} derived in Appendix \ref{doublepole_sec} we find
\begin{equation} \label{deltakappadoublepolelambda}
	\frac{\delta \kappa}{\overline\kappa} \simeq \frac{2 m \epsilon_0 Z\alpha}{n}\,\left[ \frac{ \zeta_s - \hat\lambda_0 - Z\alpha\,m\epsilon_0\,(2+\zeta_s-\hat\lambda_0) }{\zeta_s + \hat\lambda_0 }\right]
\end{equation}
for $\kappa$ found perturbatively near the IR fixed point using \pref{RGlambda}. To leading order in $Z\alpha\,m\epsilon_0$ \pref{deltakappadoublepolelambda}, gives
\be \label{deltakappaKG}
 \frac{\delta \kappa}{\overline\kappa} \simeq  \frac{2m\epsilon_0 Z \alpha}{n }  \left( \frac{\zeta_s- \hat\lambda_0}{\zeta_s+\hat\lambda_0} \right)  \,,
\ee
where we drop all subdominant powers of $Z\alpha$ and as before $\hat \lambda = 1 + mh/\pi \epsilon = 1 + h_\KG/(2\pi \epsilon)$. The fractional energy shift of the $s$-wave states (using non-relativistic kinematics, as appropriate for the leading order effect) is then
\be
  \frac{\delta E_n}{E_n} \simeq 2  \left( \frac{ \delta \kappa}{\overline\kappa} \right) \simeq \frac{4m\epsilon_0 Z \alpha}{n  } \left( \frac{\zeta_s- \hat\lambda_0}{\zeta_s+\hat\lambda_0} \right) \,,
\ee 
and so using $E_n \simeq -(Z\alpha)^2 m/(2n^2)$ we have the main result:
\be \label{Eshift}
  \delta E_n \simeq - 2m^2\epsilon_0  \left( \frac{Z \alpha}{n} \right)^3  \left( \frac{\zeta_s- \hat\lambda_0}{\zeta_s+\hat\lambda_0} \right)  \simeq 2m^2 y\epsilon_\star  \left( \frac{Z \alpha}{n} \right)^3   \,.
\ee 
Here the last equality specializes to the case $\hat \lambda_0 \to 0$ (if $y = -1$) or to $\hat\lambda_0 \to \infty$ (if $y = +1$).  For instance, if $h_0 = 0$ at $\epsilon = \epsilon_0$, then $y = +1$ and $\hat\lambda_0 = +1$ leading to $\zeta_s - \hat \lambda_0 \simeq - 2 (Z\alpha)^2$ and so
\be
  \delta E_n \simeq + 2\left[  \frac{(Z \alpha)^5}{n^3} \right] \epsilon_0 m^2  \,.
\ee 

What is noteworthy about these expressions is that they are {\em linear} in the UV scale $\epsilon_0$, precisely as was the Deser formula, above. This linearity differs from the usual assessment of finite-size effects, such as for the effects in atoms of the finite size of the nucleus, which arise quadratically in the charge-radius of the nucleus. The Deser formula is also of practical value since trading $\epsilon_\star$ (or $\epsilon_0$) for the contact-interaction scattering length, $a_s$, again leads to  \pref{dEdeser}.\footnote{In the appendix we examine  a toy model of nuclear charge, to develop intuition as to why the boundary conditions should care about the mass of the particle orbiting the nucleus.}

It is useful to quote these results in a more transparent way. For these purposes recall that a potential of the form $V = h_{\rm eff} \, \delta^3(x)$ naively shifts atomic energy levels by an amount 
\be \label{dEvsheff}
 \delta E_n = h_{\rm eff} |\psi^{(c)}_n(0)|^2 \simeq \frac{h_{\rm eff}}{\pi} \left( \frac{Z\alpha m}{n} \right)^3 \,.
\ee
This corresponds to an operator
\begin{equation} \label{hefflambda}
 h_{\text{eff}} =  -\frac{2\pi\epsilon_0}{m}\,\left( \frac{ \zeta_s - \hat\lambda_0}{\zeta_s + \hat\lambda_0 } \right)+ 2 \pi \,Z\alpha\,\epsilon_0^2\,\left( \frac{2+\zeta_s-\hat\lambda_0 }{\zeta_s + \hat\lambda_0 }\right)
\end{equation}
using \eqref{deltakappadoublepolelambda}. A given charge distribution of the nucleus parametrizes the boundary condition as
\begin{equation}
 \hat\lambda_0 = \hat\lambda_{(0)} + \hat\lambda_{(1)} (k\epsilon_0)^2 + \mathcal{O}(k\epsilon_0)^4
\end{equation}
where $k$ is the momentum inside the nucleus. Generically, in the ultra-relativistic limit $m\epsilon_0\ll Z\alpha$ the first term in \eqref{hefflambda} will dominate while the second term or a combination of the two terms dominates in the non-relativistic limit $m\epsilon_0\gg Z\alpha$ and yields $h_{\rm eff} = \frac{2\pi}{3} \, Z \alpha\, r_p^2$. Hence, interpreting \eqref{hefflambda} as predictions for an `effective' charge radius as a function of orbiting particle mass, $m$, the value of $h_{\text{eff}}$ strongly depends on how $m\epsilon_0$ compares to $Z\alpha$ and is not simply given by $\frac{2\pi}{3} \, Z \alpha\, r_p^2$. We have demonstrated this point in Figure \ref{heff_fig} below.

\begin{figure}
\centering
\includegraphics[width=0.88\textwidth]{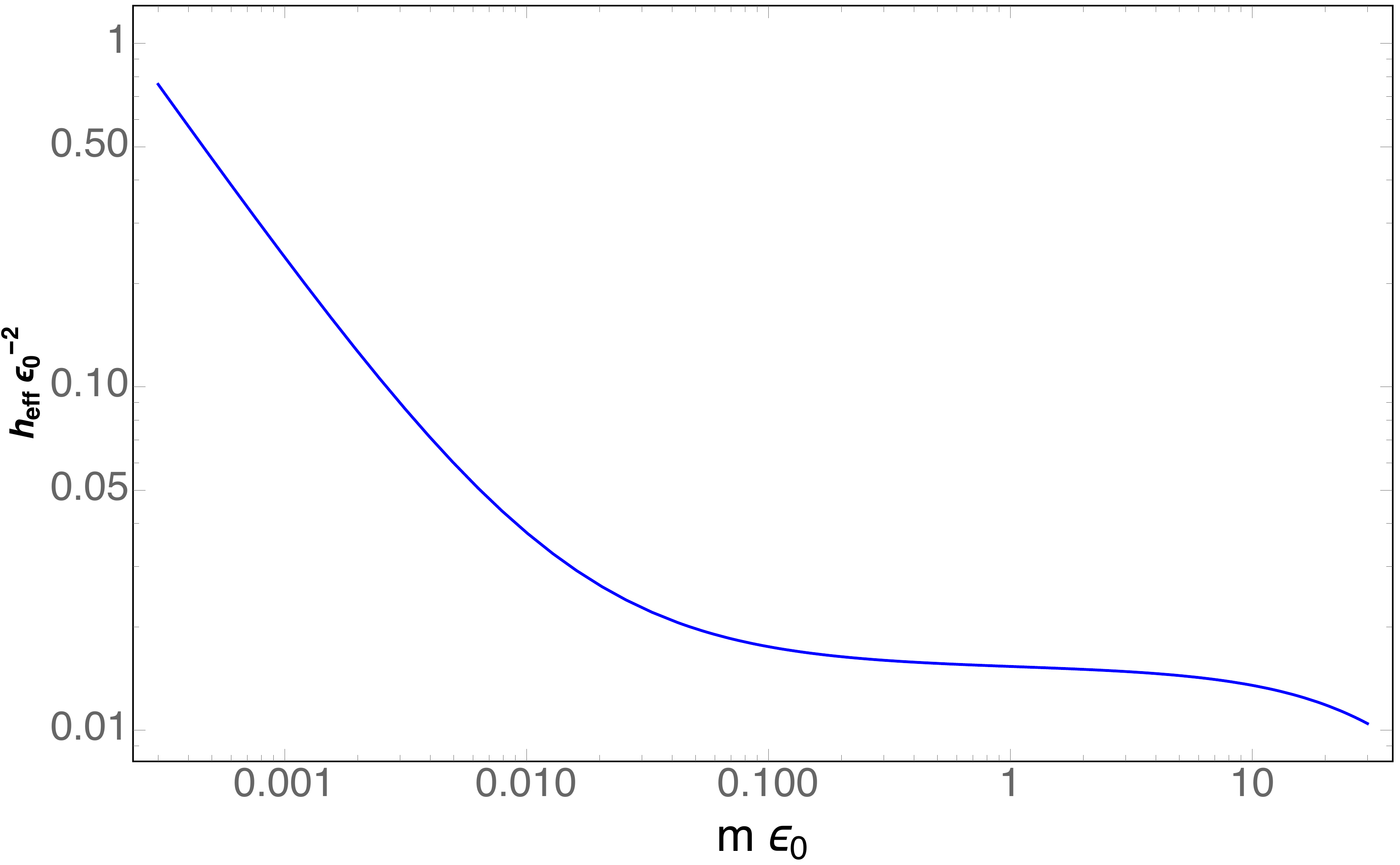}
\caption{$h_{\text{eff}}/\epsilon_0^2$ as a function of $m\epsilon_0$. The coefficients are taken to be $\hat\lambda_{(0)} = 1$ and $\hat\lambda_{(1)} = \frac43$ as appropriate for a spherical surface charge distribution discussed in Appendix \ref{chargedshell_sec}. The dispersion relation for $k$ is given in \eqref{KGdisp}.}
\label{heff_fig}
\end{figure}

\section*{Acknowledgements}

We thank Brian Batell, Richard Hill, Ted Jacobson, Friederike Metz, Sasha Penin, Maxim Pospelov, Michael Trott and Itay Yavin for helpful discussions and Ross Diener, Leo van Nierop and Claudia de Rham for their help in understanding singular fields and classical renormalization. We are grateful to Marko Horbatsch and Henry Lamm for their careful reading of the manuscript, including the catching of several errors. This research was supported in part by funds from the Natural Sciences and Engineering Research Council (NSERC) of Canada and by a postdoctoral fellowship from the National Science Foundation of Belgium (FWO). Research at the Perimeter Institute is supported in part by the Government of Canada through Industry Canada, and by the Province of Ontario through the Ministry of Research and Information (MRI).  

\appendix

\section{Matching to a simplistic nuclear model}

In this appendix we describe several simple toy models of a nuclear charge distribution, with the goal of making more explicit how $\lambda = 2mh$ should be expected to depend on $m$. 

We examine two distributions: one where all of the nuclear charge is located at the nuclear surface, $r = R$, and one where the charge is uniformly distributed throughout the nucleus, $r \le R$. We show both predict $\lambda \propto m$ (and so $h$ to be roughly $m$-independent) when computed within the non-relativistic Schr\"odinger regime, but both also predict $\lambda$ to be $m$-independent (and so $h \propto 1/m$) when examined in the regime where the orbiting particle would be relativistic at the nuclear surface.

We always demand $R$ to be much smaller than the Bohr radius, which implies $R \ll 1/(Z\alpha \,m)$. In the Schr\"odinger analysis we also demand $m \gg 1/R$ and so its range of validity is for the window 
\be \label{mRZa}
  \frac{1}{Z\alpha } \gg mR \gg 1  \,,
\ee
which is non-empty because $Z \alpha \ll 1$. The relativistic analysis requires only the first of these inequalities and so assumes only $mR Z \alpha \ll 1$.

\subsection{Spherical surface-charge distribution} \label{chargedshell_sec}

The simplest (but least realistic) distribution assumes that the charge is concentrated in an infinitely thin sphere at $r=R$:
\be
 \rho =   \sigma_0\, \delta(r-R)  
\ee
where the constant charge per unit area, $\sigma_0$, is related to the total charge by  $\sigma_0 = Ze/(4\pi R^2)$. In this case, the electrostatic potential is  
\be
 A^0 = \left\{ \begin{array}{cc} Z e /(4\pi R)  &  \hbox{for $r \le R$} \\ Z e /(4\pi r) & \hbox{for $r > R$}   \end{array}  \right. \,,
\ee
which is chosen to be continuous at $r=R$ with the external Coulomb potential.

\subsubsection*{Schr\"odinger formulation}

Let us discuss the $s$-wave solution with this potential. Outside the charged sphere ($r > R$) it is the Schr\"odinger solution, $\psi_{\rm out}(r)$, for the Coulomb problem, though without imposing regularity at the origin. We denote the energy of the state by $E$ and determine this by matching the solution to the one found for $r \le R$. 

Inside the charge sphere $(r<R)$ the wavefunction is that of a free particle, for which we choose regularity at the origin (because there is no source located there). This leads to the interior solution
\begin{equation} \label{freeswave}
 \psi_{\rm in}(r) = C_{\rm in} \frac{\sin(k r)}{r}\,,
\end{equation}
with $k$  given in terms of $E$ by
\begin{equation} \label{kepsA}
 k^2 = 2 m \left(  E + eA^0 \right)= 2 m \left(  E + \frac{Z \alpha}{R}  \right) \simeq \frac{2mZ\alpha}{R} \,,
\end{equation}
where the last, approximate, equality uses the condition $R \ll 1/(Z\alpha \, m)$ to infer $Z\alpha/R \gg |E|$, since in the ground state $|E| \simeq \frac12 (Z\alpha)^2 m$. 

The wave function and its derivative must be continuous across $r=R$, and matching $\psi_{\rm in}(R) = \psi_{\rm out}(R)$ relates the overall normalization constants of $\psi_{\rm in}(r)$ and $\psi_{\rm out}(r)$. For the present purposes it is the matching of the derivatives that is more interesting, which can be written as
\begin{equation}
 \frac{\psi_{\rm out}'(R)}{\psi_{\rm out}(R)} = \frac{\psi_{\rm in}'(R)}{\psi_{\rm in}(R)} = k \cot(k R) - \frac{1}{R}\,,
\end{equation}
showing a possible underlying origin of the nontrivial boundary condition entertained in the main text at small $r$.

On the other hand, recall that outside the nucleus for sufficiently small $\kappa \epsilon$ the wave-function $\psi_{\rm out}$ satisfies
\begin{equation}
 \frac{\psi_{\rm out}'(\epsilon)}{\psi_{\rm out}(\epsilon)} = \frac{\lambda(\epsilon)}{4 \pi \epsilon^2} = \frac{\hat \lambda(\epsilon) - 1}{2 \epsilon}\,,
\end{equation}
using $\hat \lambda = \lambda / (2 \pi \epsilon) +1$. Applying this to $\epsilon \to R$, the logarithmic derivative of the wavefunction of the interior $\psi_{\rm in}(R)$ effectively fixes the function $\hat \lambda(R)$. The $m$-dependence and other properties of $\hat \lambda(R)$ in the external theory can be directly related to the properties of the source through this matching condition
\begin{equation}
 \frac{\psi_{\rm in}'(R)}{\psi_{\rm in}(R)} = \frac{\hat \lambda(R) - 1}{2 R}\,.
\end{equation}

For our toy model  we find in this way
\begin{equation} \label{lambdaeq}
 \lambda(R) = 4 \pi R^2 \; \frac{\psi_{\rm out}'(R)}{\psi_{\rm out}(R)} = 4 \pi k R^2 \left[ \cot(k R) - \frac{1}{k R} \right]  \simeq - \frac{4\pi k^2R^3}{3} \simeq - \frac{8 \pi}{3} m R^2 Z \alpha\,,
\end{equation}
where we use \pref{kepsA} to infer $(kR)^2 \simeq 2mR Z\alpha \ll 1$, with this last inequality following from \pref{mRZa}. We see this model predicts 
\be
 h(R) = \frac{\lambda(R)}{2m} \simeq - \frac{4 \pi R^2 Z \alpha}{3} \,,
\ee
which is the same for any particle (independent of their mass) at the matching scale $R$.

\subsubsection*{Klein-Gordon formulation}

We can describe the same distribution using the KG equation, in order to treat the regime where $m \le 1/R$. We still require $R$ to be much smaller than the Bohr radius, and so continue to require $mRZ \alpha \ll 1$. To do so we compute the matrix element, $\psi(x) = \langle 0 | \Psi(x) | n \rangle$, where $\Psi$ is the KG field and $|n\rangle$ is an atomic meson state. 

For $s$-wave solutions with energy $\omega$ this function $\psi(r)$ solves the KG equation, with solutions still given by \pref{freeswave} but dispersion relation giving $k$ now being
\begin{equation}
 \left(\omega + \frac{Z \alpha}{R} \right)^2 - k^2 = m^2\,.\label{KGdisp}
\end{equation}
This reduces to the non-relativistic \Sch dispersion relation  for $\omega = m + E$ with, as before, $E \simeq -\frac12 (Z\alpha)^2 m$ for the ground state. In the regime $Z\alpha/R \gg \omega \simeq m$ this dispersion relation can be approximated as
\begin{equation}
 (kR)^2 \simeq  (Z \alpha)^2 \ll 1\,.
\end{equation}

Again expanding \pref{lambdaeq} for small $kR$ we get
\begin{equation}
 \lambda= 4\pi R \Bigl[ kR \cot(kR) - 1\Bigr] \simeq - \frac{4\pi k^2R^3}{3} \simeq- \frac{4 \pi}{3} (Z \alpha)^2 R\,,
\end{equation}
which shows that $\lambda$ in this regime is independent of $m$ (as must also be the KG source coupling $h_\KG = \lambda$). The equivalent \Sch coupling therefore becomes
\be
 h = \frac{h_\KG}{2m} = \frac{\lambda}{2m} = - \frac{2\pi (Z\alpha)^2R}{3m} \,,
\ee
which varies inversely with $m$.

\subsection{Constant charge distribution}

A slightly more realistic choice is a constant charge distribution:
\be
 \rho = \left\{ \begin{array}{cc} \rho_0 & \hbox{if $r < R$}  \\ 0 & \hbox{if $r > R$}    \end{array} \right.
\ee
where the constant $\rho_0$ is related to the total charge by $\rho_0 = 3Ze/(4\pi R^3)$.  In this case the electrostatic potential $\varphi = A^0$ satisfies 
\be
 A^0 = + \frac{Ze}{4\pi R} - \frac{\rho_0}{6} \Bigl( r^2 - R^2\Bigr)   = -\frac{Ze}{8\pi R} \left( \frac{r^2}{R^2} - 3 \right) \,,
\ee
where the integration constants ensure $A^0$ is nonsingular at $r = 0$ and is continuous with the external Coulomb potential at $r = R$.

\subsubsection*{Schr\"odinger formulation}

$s$-wave solutions to the \Sch equation with this potential satisfy
\be \label{swaveconst}
     \frac{1}{r^2} \, \partial_r \Bigl( r^2 \partial_r \psi \Bigr) = -\left( E + e A^0 \right) \psi
   = -\left( V_0 -  V_2 \,r^2 \right) \psi \,,
\ee
where 
\be
  V_0  = 2m\left( E + \frac{3Z\alpha}{2 R} \right) \quad \hbox{and} \quad 
  V_2   = \frac{mZ \alpha}{ R^3}  \,.
\ee

Eq.~\pref{archetype} has as its general solution
\be
 \psi(x) = \frac{1}{x} e^{- x^2/2} \Bigl[ C_+ \psi_+(x) + C_- \psi_-(x) \Bigr]  \,,
\ee
where $\psi_{\pm}(x)$ are a pair of basis solutions that can be written in terms of confluent hypergeometric functions and the dimensionless coordinate is $x = \mu \, r$ where 
\be
 \mu^4 = V_2 = \frac{m Z \alpha}{R^3} \,.
\ee
Since \pref{swaveconst} is invariant under $r \to -r$ we may choose $\psi_\pm(-r) = \pm \psi(r)$, in which case $C_+ = 0$ is required for regularity at $x=0$, and $\psi_-(x)$ is ultimately a series in powers of $x^2 = \mu^2 r^2$, given explicitly by
\be
 \psi_-(x) = x -2 (\nu-1)  \, \frac{x^3}{3!} +  \cdots  \,,
\ee
where
\be
 \nu := \frac{V_0- \mu^2}{2\mu^2} = \left( E R + \frac{3Z\alpha}{2} \right)\sqrt{\frac{mR}{Z \alpha}}  - \frac12 
  \simeq \frac12 \left( 3 \sqrt{mRZ\alpha} - 1 \right) \simeq - \frac12\,,
\ee
which simplifies using \pref{mRZa}. Because $(\mu r)^2 \le (\mu R)^2 = mRZ\alpha$ eq.~\pref{mRZa} also says that the regime of interest is small $x$ for which $\psi(r) \simeq C_{-} \left[ 1 + \frac12\, \cC_2 x^2 + \frac14\, \cC_4 x^4 + \cdots \right]$ with $\cC_2(\nu) = - \frac{1}{3} \left(\nu+\frac12\right) \simeq 0$. Therefore $\partial_r \ln \psi_{\rm in}(r=R) \simeq \cC_4 \mu^4R^3$, and so
\be
 \lambda(R) = 4 \pi R^2 \, \partial_r \ln \psi_{\rm in}(R) \simeq 4\pi \cC_4 \, \mu^4 R^5  \simeq 4\pi \cC_4 \, mR^2 Z \alpha \,.
\ee
This again predicts $\lambda \propto m$ and so $h = \lambda/2m$ independent of $m$.

\subsubsection*{Klein-Gordon formulation}

The KG equation to be solved with this potential is in this case
\be 
    \nabla^2  \psi = \frac{1}{r^2} \, \partial_r \Bigl( r^2 \partial_r \psi \Bigr) = \left[m^2 - \left( \omega + e A^0 \right)^2 \right] \psi
   = -\left( V_0 +  V_2 \,r^2 + V_4 r^4 \right) \psi \,,
\ee
where  we  define the constants
\bea
  V_0 &=&  \left( \omega + \frac{3Z\alpha}{2 R}  \right)^2 - m^2 \simeq \frac94  \left( \frac{Z\alpha}{ R}  \right)^2  \nn\\ 
  V_2  &=& -\left( \omega + \frac{3Z\alpha}{2R}  \right)  \frac{Z \alpha}{R^3} \simeq -\frac3{2R^2}  \left( \frac{Z\alpha}{ R}  \right)^2 \\
  V_4 &=& \left( \frac{Z\alpha}{2R^3} \right)^2 = \frac1{4R^4}  \left( \frac{Z\alpha}{ R}  \right)^2\,, \nn
\eea
and we focus on the regime $Z\alpha/R \gg \omega \simeq m$. The above reduces to the \Sch result when $\omega = m + E$ and we take $m \gg 1/R$ and $E \simeq (Z\alpha)^2 m$ as before.

Although not simply solvable, its dependence on scales is made explicit by changing coordinates to $z = r/R$ and multiplying the equation through by $R^2$, giving
\be \label{archetype}
 \psi'' + \frac{2\psi'}{z} + \Bigl( A + B z^2 + C z^4 \Bigr) \psi = 0 \,,
\ee
with $\psi' := \partial_z \psi$ and
\be
 A = R^2 V_0 \simeq \frac94  \left( Z\alpha \right)^2  \,, \quad
 B =R^4 V_2  \simeq \frac32  \left( Z\alpha \right)^2 \quad \hbox{and} \quad
 C = R^6 V_4 \simeq \frac14  \left( Z\alpha \right)^2\,.
\ee
Our interest is $r < R$ and so $z < 1$.

To evaluate $\lambda$ we can then approximate $\psi_{\rm in}(z) \simeq f[z^2,(Z\alpha)^2]$ with $f$ going over to spherical Bessel functions as $Z \alpha \to 0$. Therefore $\partial_r \ln \psi_{\rm in}(r=R)$ becomes
\be
 \lambda = 4 \pi R^2 \partial_r \ln \psi_{\rm in}(R) \simeq 4\pi \widetilde\cC \, R  \,,
\ee
where $\widetilde\cC$ is an order $(Z \alpha)^2$ number obtained by evaluating $f'/f$ at argument $z = 1$. This regime again predicts $\lambda = h_\KG$ to be independent of $m$, and so $h \propto R/m$.

\section{Gamma Function approximations when $\zeta_s \lesssim 1$} \label{doublepole_sec}

\begin{figure}
\includegraphics[width=0.88\textwidth]{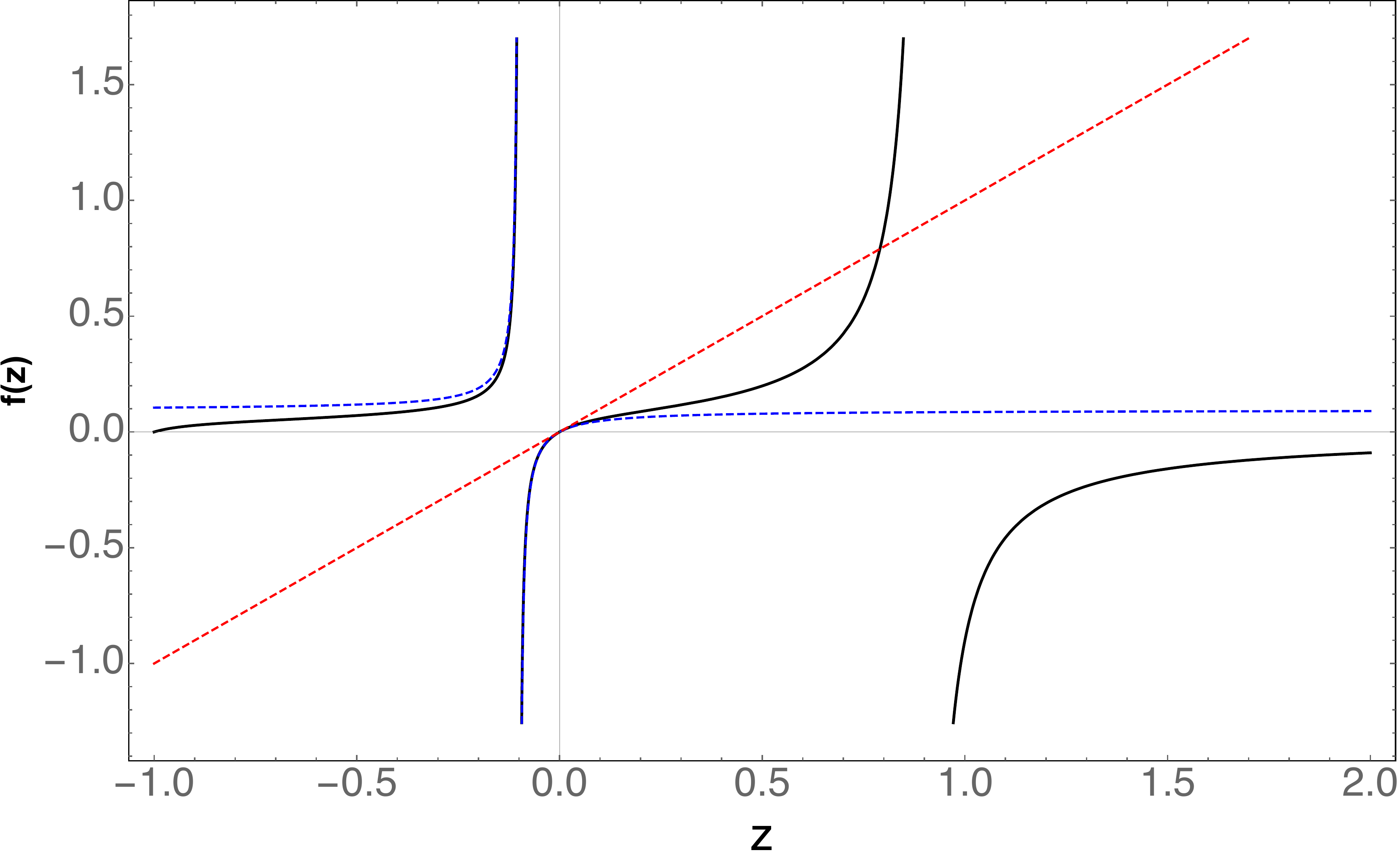}
\caption{$f(z)$ which resembles the RHS of \eqref{C-C+App}. We are interested in a precise approximation of this function close to the origin $z=0$. The black solid line is $f(z) = \Gamma(z-a)/\Gamma(z)/\Gamma(-a)$, the blue dashed line $f(z) = -z/(z-a+1)/\Gamma(-a)$ (double pole approximation) and the red dashed line $f(z) = z$ (single pole approximation). The chosen numerical value is $a=0.9$.}
\label{GGGG_fig}
\end{figure}

The ratio $C_-/C_+$ is given by a ratio of Gamma functions \eqref{inftybc}
\begin{equation} \label{C-C+App}
 \frac{C_-}{C_+} =  \frac{\Gamma(\zeta_s) \Gamma\left[ \frac12 \left( - \frac{w}{\kappa} +1 - \zeta_s\right)\right]}{\Gamma(-\zeta_s) \Gamma\left[ \frac12 \left( - \frac{w}{\kappa} +1 + \zeta\right)\right]} \,.
\end{equation}
We can rewrite this expression as
\begin{equation}
 \frac{C_-}{C_+} =  \frac{\Gamma(\zeta_s) \Gamma(z+1-\zeta_s-(N+1))}{\Gamma(-\zeta_s) \Gamma(z-N)} \,.
\end{equation}
where
\begin{equation}
 z \equiv \frac12 (-\frac{w}{\kappa}+1+\zeta_s) + N \ll 1\,,
\end{equation}
since $\kappa$ is close to the Bohr energy \eqref{kappaquant}. If $|\zeta_s - 1| \ll 1$, as for instance in the Klein-Gordon Coulomb problem, both $\Gamma$-functions depending on $z$ are in the vicinity of a pole, see Figure \ref{GGGG_fig}. This makes it necessary to approximate both $\Gamma$-functions by their respective poles whereas \eqref{singlepole} is sufficient if $\zeta_s$ is not close to one.

Using $\Gamma(z-N) \simeq (-1)^N / (N!\,z)$ for $z\ll1$ we can then make the approximations
\begin{align}
 \begin{aligned}
  \Gamma(z-N) &\simeq \frac{(-1)^N}{N!\,z}\,,\\
  \Gamma(z+1-\zeta_s-(N+1)) &\simeq \frac{(-1)^{N+1}}{(N+1)!\,(z+1-\zeta_s)}\,,\\
  \Gamma(\zeta_s) &\simeq 1\,,\\
  \Gamma(-\zeta_s) &\simeq \frac{-1}{1-\zeta_s}\,.
 \end{aligned}
\end{align}
With these approximations \eqref{inftybc} becomes
\begin{equation}
 \frac{C_-}{C_+} \simeq \frac{(1-\zeta_s)z}{n\,(z+1-\zeta_s)}\,.
\end{equation}
With $z \simeq n \delta \kappa / \bar \kappa$ we then find
\begin{equation}\label{deltakappadoublepole}
 \frac{\delta \kappa}{\bar \kappa} \simeq \frac{(1-\zeta_s)C_-/C_+}{(1-\zeta_s)-n\,C_-/C_+}\,.
\end{equation}

\end{document}